\documentclass[conf]{new-aiaa}
\usepackage[utf8]{inputenc}

\usepackage{graphicx,color}
\usepackage{amsmath}
\usepackage[version=4]{mhchem}
\usepackage{siunitx}
\usepackage{longtable,tabularx}
\usepackage{subcaption}
\usepackage{hyperref}

\newcommand{\eqnref}[1]{Eq.~(\ref{#1})}
\newcommand{\figref}[1]{Fig.~\ref{#1}}
\newcommand{\tabref}[1]{Tab.~\ref{#1}}
\newcommand{\algref}[1]{Alg.~\ref{#1}}

\usepackage{bm}

\usepackage{optidef} % to write optimization problems
\usepackage[ruled,vlined]{algorithm2e}  % to write algorithms
\usepackage{mathrsfs}   % to write cursive letters
\usepackage{changepage}   % to prevent overfull box
\usepackage{float} 
\usepackage{dsfont}

\usepackage{comment}

\title{Bayesian optimization for mixed variables using an adaptive dimension reduction process: applications to aircraft design}

\author{ P. Saves\footnote{PhD Student, Information Processing and Systems Department \& Complexes Systems Engineering Department, paul.saves@isae-supaero.fr}, E. Nguyen Van\footnote{Research Engineer, Information Processing and Systems 
Department, eric.nguyen\_van@onera.fr}, N. Bartoli \footnote{Senior researcher, Information Processing and Systems Department, nathalie.bartoli@onera.fr, AIAA MDO TC Member.}, T. Lefebvre\footnote{Research Engineer, Information Processing and Systems 
Department, thierry.lefebvre@onera.fr, AIAA Member.}, C. David\footnote{Research Engineer, Information Processing and Systems 
Department, christophe.david@onera.fr},  S. Defoort\footnote{Research Engineer, Information Processing and Systems 
Department, sebastien.defoort@onera.fr}}
\affil{ONERA, DTIS, Universit\'e de Toulouse, Toulouse, France}
\author{Y. Diouane\footnote{Professor, Complexes Systems Engineering Department, youssef.diouane@isae-supaero.fr}}
\affil{ISAE-SUPAERO, Université de Toulouse, Toulouse, France}

\author{J. Morlier\footnote{Professor, Structural Mechanics, joseph.morlier@isae-supaero.fr, AIAA Member.}}
\affil{ICA, Universit\'e de Toulouse, ISAE-SUPAERO, MINES ALBI, UPS, INSA, CNRS, Toulouse, France}

\begin{document}

\maketitle
%\tableofcontents
\begin{abstract}
Multidisciplinary design optimization methods aim at adapting numerical optimization techniques to the design of engineering systems involving multiple disciplines. In this context, a large number of mixed continuous, integer and categorical variables might arise during the optimization process and practical applications involve a large number of design variables.
%Among MDO architectures, various ones are considering the resolution of the Multidisciplinary Design Analysis (MDA). In our study, the system of interest being an aircraft, the resolution of the MDA  will be provided by the Future Aircraft Sizing Tool with Overall Aircraft Design (FAST-OAD), a point mass approach that estimates the required fuel and energy consumption for a given set of top-level aircraft requirements. 
Recently, there has been a growing interest in mixed variables constrained Bayesian optimization but most existing approaches severely increase the number of the hyperparameters related to the surrogate model. 
In this paper, we address this issue by constructing surrogate models using less hyperparameters. The reduction process is  based on the partial least squares method. An adaptive procedure for choosing the number of hyperparameters is proposed. The performance of the proposed approach is confirmed on analytical tests as well as two real applications related to aircraft design. A significant improvement is obtained compared to genetic algorithms.

\end{abstract}

%%%%%%%%%%%%%%%%%%%%%%%%%%%%%%%%%%%%%%%%%%%%%%%%%%%%%%%%%%%%%%%%%%%%%%

% Class automatically runs \maketitle at the begin of document, producing
%
% Title, authors and affiliations, abstract and keywords

%%%%%%%%%%%%%%%%%%%%%%%%%%%%%%%%%%%%%%%%%%%%%%%%%%%%%%%%%%%%%%%%%%%%%%

%%%%%%%
%\tableofcontents
%%%%%%

\section{Introduction}
\label{sec:intro}
New aircraft configurations with a lower footprint on the environment (also known as Eco-aircraft design) have seen a resurgence of interest \cite{Eco-material,SEGO-UTB-Bombardier}. In this context, one targets to minimize the footprint on the environment of the aircraft using a \textit{Multidisciplinary Design Analysis} (MDA)~\cite{Lambe2011,Lambe2012,Lambe2013}. The process of finding the best configuration, known as \textit{Multidisciplinary Design Optimization} (MDO), is formulated as a minimization constrained problem where the objective and the constraint functions are typically expensive-to-evaluate and their derivatives are not available.% Moreover as the number of design variables could increase drastically, some specific strategies have to be investigated to handle this complexity. For instance the Partial Least Squares technique has been proposed~\cite{Bouhlel2016improving,Bouhlel2016_MPE} and will be considered in an adaptive process. \\

In the context of aircraft design, the MDO process generally involves mixed continuous-categorical design variables. For instance, the size of aircraft structural parts can be described using continuous variables; in case of thin-sheet stiffened sizing, they represent panel thicknesses and stiffening cross-sectional areas. The set of discrete variables can encompass design variables such as the number of panels, the list of cross sectional areas or the material choices. Thus, the regarded optimization problem is of the following form:
\begin{equation}
   \min_{w=(x,z,c) \in \Omega \times S \times \mathbb{F}^l}  \left\{ f(w) ~~\mbox{s.t.}~~ g(w)\leq 0 ~\mbox{and}~ h(w)=0 \right\}
    \label{eq:opt_prob}
\end{equation}
$f:\mathbb{R}^n \times  \mathbb{Z}^m \times \mathbb{F}^l \mapsto \mathbb{R}$ is the objective function, $g:\mathbb{R}^n \times  \mathbb{Z}^m \times \mathbb{F}^l \mapsto \mathbb{R}^m$ gives the inequality constraints, and $h:\mathbb{R}^n \times  \mathbb{Z}^m \times \mathbb{F}^l \mapsto \mathbb{R}^p$ returns the equality constraints.
 $\Omega \subset \mathbb{R}^n$ represents the bounded continuous design set for the $n$ continuous variables.  $S \subset \mathbb{Z}^m$ represents the bounded integer set where $L_1, . . . , L_m$ are the numbers of levels of the $m$ quantitative integer variables on which we can define an order relation and $ \mathbb{F}^l = \{1, \ldots, L_1\} \times \{1, \ldots, L_2\} \times  \ldots \times \{1, \ldots, L_l\}$ is the design space for $l$ categorical qualitative variables with their respective  $L_1, . . . , L_l$ levels.
The functions $f$, $g$, and $h$ are typically expensive-to-evaluate simulations with no exploitable derivative information.

When only continuous design variables are optimized (i.e., $\mathbb{F}^c$ is reduced to a single categorical choice), \textit{Bayesian optimization} (BO) is shown to be powerful strategy for solving problem~(\ref{eq:opt_prob})~\cite{Mockus_book}. BO uses \textit{Gaussian processes} (GPs)~\cite{Jones98,Mockus,Rasmussen,forrester,Sasena} (also denoted by Kriging~\cite{krige1951statistical}) to define response surface models, and the sequential enrichment is performed by maximizing a given acquisition function~\cite{Jones98}. The latter is meant to model a compromise between exploration of new zones in the design space and exploitation (i.e. minimization) of the GPs. 
For general mixed integer problems, several modeling strategies to build GPs have been proposed~\cite{Pelamatti, Roustant18}. Based on these GP models, a trade-off acquisition function was adapted for optimization~\cite{CAT-EGO}. Some other modelling strategies consist in computing a continuous model for each category~\cite{AMIEGO19}, either by continuously relaxing the design variables~\cite{GMHL}, by using  a multi-armed bandit strategy to handle the categorical choices~\cite{Bandit-BO} or by considering a Gower distance to model simultaneously the proximity over categorical and continuous variables~\cite{Gower}.
Recently, a continuous relaxation BO based method~\cite{well-adapted_cont} to tackle mixed integer variables has been shown to solve efficiently expensive-to-evaluate optimization problems. In fact, using continuous relaxation within BO leads to better results. However, the relaxation of the categorical design variables increases the number of the hyperparameters needed (to be tuned) associated with the GP model. This in particular constrained the method in~\cite{well-adapted_cont} to be used only for small dimensional optimization problems. Since, the construction of the GP model may not be scalable to practical applications involving a large number of mixed variables, some specific strategies have to be investigated to handle this complexity: the Kriging model with Partial Least Squares (KPLS) is one of the most used technique~\cite{Bouhlel2016improving,Bouhlel2016_MPE}.

In this work, we show how to reduce the computational cost related to the construction of the mixed categorical GP model as proposed in~\cite{GMHL}. Our proposed approach allows in particular to solve higher dimensional mixed integer MDO problems. The method relies on combining continuous relaxation and the use of KPLS. Inspired by the Wold's R criterion
~\cite{PLS_Li_2002,PRESS,WoldPLS}, we propose also an adaptive mechanism for choosing the number of hyperparameter to use within KPLS. The good potential of the proposed approach is first showed over a set of analytical test cases. 

The performance of the proposed approach is then confirmed on two MDO applications from the FAST-OAD framework~\cite{David_2021}: ``\texttt{CERAS}'' and ``\texttt{DRAGON}''. FAST-OAD is an open-source Python framework that provides a flexible way to build and solve the Overall Aircraft Design problems by assembling discipline models from various sources: it currently comes with some bundled, quick and simple, models dedicated to commercial aircraft. In this paper, FAST-OAD will resolve an MDA problem that mainly: (a) sizes the geometry of main aircraft components, (b) computes mass and centers of gravity of aircraft parts, (c)  estimates the aerodynamics and propulsion along the computed mission, and  (d) returns the fuel consumption related to the mission. These estimated quantities will be used to define the objective and the constraints of our two optimization problems.

The outline of the paper is as follows.
In Section~\ref{sec:BO}, a detailed review of the Bayesian optimization framework is given.
The continuous relaxation as well as the use of the KPLS technique are detailed in Section~\ref{sec:met}.
Section~\ref{sec:res} presents our academical tests as well as the obtained results on the two regarded MDO test cases.
Conclusions and perspectives are finally drawn in Section~\ref{sec:conclu}.

\section{Continuous constrained Bayesian optimization}
\label{sec:BO}
%For mixed categorical integer Bayesian optimization, several heuristics have been proposed that randomly select points over mixed variables by defining a neighbourhood-like metric~\cite{SO-MI, Abramson2002,CAT-EGO}. However, genetic and evolutionary algorithms have been shown to perform well~\cite{NSGA2,Pelamatti}. 

In this section, we will only consider that all the design variables are continuous in problem \eqref{eq:opt_prob}. Namely, in this section, the design space will be restricted to $\Omega \subset \mathbb{R}^n$; hence $w=x$ in the optimization problem \eqref{eq:opt_prob}.  %as the principle of Bayesian optimization remains the same in this case. 
In the context of unconstrained optimization, the Efficient Global Optimization (EGO) framework~\cite{Jones98} suggests to use the GP model to find the global minimum of an expensive-to-evaluate black-box function (based on the preliminary results of Mo{\v{c}}kus~\cite{Mockus}). In general,
a GP is used to fit a response surface model from an initial set of points known as the Design of Experiments (DoE)~\cite{forrester,KIM, Rasmussen}.
The GP provides a mean response hypersurface as well as a pointwise estimation of its variance. Thereafter, we will consider that our unknown black-box objective function $f$ is a realization of an underlying distribution of mean $\hat{f}$  and of standard deviation $s^f$ such that $f(.) \sim GP(\hat{f}(.), [s^f]^2(.))$. \\
Let $n_t$ be the number of already evaluated points in $\mathbb{R}^{n} $ of the deterministic function $f$  and $ \forall i \in \{1,..,n_t\}$, let ${x}^{i} =(x_1^{i},...,x_{n}^{i}) \in \mathbb{R}^{n}$ be the $i^{th}$ point with its respective $n$ continuous variable values. The stochastic model~\cite{GP14} writes as: $f(x) = \mu ({x})+\epsilon \in \mathbb{R} $ with $\epsilon$ the error term between $f$ and the model approximation $\mu(x)$. The errors terms are considered $\textit{iid}$ of variance $\sigma^2$.\\
Let $R$ be the error correlation matrix between the inputs points $R_{ij}=Corr(\epsilon(x^i),\epsilon(x^j))$. The correlation function $Corr$ is computed using a kernel function $k$ that relies on $n$ hyperparameters $\theta$ estimated typically using maximum likelihood estimator (MLE)~\cite{MLE}: $Corr(.,.)=k(.,.,\theta)$. 
Let $r_i(x^*) = Corr(\epsilon(x^*),\epsilon(x^i))$ for a given $x^*$, then

\begin{equation}
\hat{f}(x^*)= \hat{\mu}^f+r(x^*)^T R^{-1}(\textbf{y}^f-\mathds{1} \hat{\mu}^f), 
\end{equation}
and
\begin{equation}
[s^f]^2(x^*)=[\hat{\sigma}^f]^2\left[1-r(x^*)^TR^{-1}r(x^*)+ \frac{(1-\mathds{1}^T R^{-1}r(x^*))^2}{\mathds{1}^T R^{-1}\mathds{1}}\right], \end{equation}
where $\hat{\mu}^f$ and $\hat{\sigma}^f$, respectively, are the MLE of $\mu$ and $\sigma$ with respect to $\theta$ given the data set $(\textbf{x},\textbf{y}^f)$~\cite{Jones98}.

Within EGO, at a given iteration $t$, a GP surrogate model, referred as Kriging, is computed based on the current DoE to approximate the black-box $f$.  Henceforth, one wants to estimate the best new point to evaluate, as it is costly, by taking into account the model information to converge as fast as possible to the real optimum of the black-box. The point that we will evaluate next is the one that gives the best improvement \textit{a priori} according to an acquisition function like the Expected Improvement (EI) defined over the model. The objective value at this new point will then be evaluated and incorporated into the next surrogate model. The hyperparameters that characterize and define the model are thus updated at each iteration until convergence.  The Bayesian optimization process is made from these GPs in an iterative manner.  

To tackle constrained Bayesian optimization, EGO was extended to Super-Efficient Global Optimization (SEGO) method~\cite{Sasena}. SEGO uses surrogate models of the constraints to give an estimation of the search space $\Omega_f$ through a given criterion in which the function $f$ is optimized. The latter was enhanced to tackle multi-modal and equality constraints with the Upper Trust Bound (UTB) criterion~\cite{SEGO-UTB,Bartoli2019,SEGO-UTB-Bombardier}. 
The acquisition function that we use is the WB2s (Watson and Barnes $2^{nd}$ criterion with scaling)~\cite{Bartoli2019} that is known to be more robust than the Expected improvement (EI) criterion, especially in high dimension~\cite{jesus2021surrogate}. WB2s can be seen as smooth version compared to the WB2~\cite{Sasena} criterion and is less multimodal compared to EI. Algorithm~\ref{alg:optim} details the SEGO optimization procedure.

\smallskip
\begin{algorithm}[H]
\SetAlgoLined
\KwResult{Solution of the problem \eqref{eq:opt_prob} over the continuous design space $\Omega$.}
{\textbf{Inputs:}}  Initial DoE $\mathscr{D}_0$ and set $t=0$. Search space $ \Omega$. 
% \newline

\While{the stopping criterion is not satisfied}{
\vspace{.2cm}
\begin{adjustwidth}{0pt}{40pt}
\begin{enumerate}
    \item Build the surrogate model of the objective function to obtain the mean and standard deviation prediction at a given point: $(\hat{f},s^f)$ from the DoE $\mathscr{D}_t$.
    \item Build the surrogate models for every constraint $(\hat{g}_i,s^g_i)$, $(\hat{h}_j,s^h_j)$ from the DoE $\mathscr{D}_t$ to compute an estimation of the search space $\Omega_f$.
    \item Construct the acquisition function $WB2s(.) = \phi\left(\hat{f}(.),s^f(.)\right)$ from the objective model. 
    \item Maximize the acquisition function $WB2s$ over $\Omega_f$: $x_{t}=  \underset{x \in \Omega_f}{\arg \max}\   WB2s(x)$. 
    \item Add $x_{t}, f(x_{t}), g(x_{t}), h(x_{t})$ to the DoE  $\mathscr{D}_{t+1}$. Increment $t$.
\end{enumerate}
\end{adjustwidth}
\smallskip

} % End while

\caption{SEGO for continuous inputs.}
\label{alg:optim}
\end{algorithm}

\smallskip

\section{Mixed categorical constrained Bayesian optimization}

\label{sec:met}

To handle mixed categorical design variables, we propose to use the continuous relaxation method that has been recently shown to be well-suited for expensive discrete problems~\cite{well-adapted_cont,GMHL}.
The main drawback of such method is that it enlarges the dimension of the design space according to the size of the categorical space. To overcome such issue, we propose to combine continuous relaxation with the well-known partial least squares procedure~\cite{Bouhlel2016improving} to reduce the number of the GP hyperparameters.\\
For mixed categorical design variables, the proposed treatment relies on continuous relaxation. The design  space  $ \Omega \times S \times \mathbb{F}^l $ is relaxed to a continuous space $ \Omega'$  constructed on the following way:

\begin{itemize}
\item $\forall i \in \{1, \ldots,m\}$, the variable $z_i$ is relaxed within its bounds and treated as continuous. 
\item   $\forall j \in \{1, \ldots,l\}$, we use a relaxed one-hot encoding~\cite{one-hot} for $c_j$ and add $L_j$ new continuous dimensions into $ \Omega'$. 
\end{itemize}
Therefore, we get, after relaxation, a new design space $\Omega'\subseteq \mathbb{R}^{n'}$ where $n'= n+m+ \sum_{j=1}^l L_j >n+m+l$. 
The nature of the variables should be respected when evaluating a point in the relaxed space so we define the inverse operator $Project$ that projects a point $  X \in \Omega'$ to its closer point $w^{X}$  in  $ \Omega \times S \times \mathbb{F}^l $. Namely, $Project$ rounds the value of an  integer variable $z_i$ to the closer value among its $L_i$ levels and, for a categorical variable $c_j$, $Project$ selects the level which corresponding dimension value is the highest.

In this work, when building the Kriging model, the error correlation will be estimated using a squared exponential (or Gaussian) correlation kernel over the relaxed design space. We denote by $X^{w_i} = Relax(w^i)$ the relaxation in $\Omega'$ of a point  $w^i \in \Omega \times S \times \mathbb{F}^l $. The mixed categorical kernel is 
\begin{equation}
k(X^{w^i},X^{w^j},\theta) = \prod\limits_{p=1}^{n'}\exp\left(-\theta_p\left(X_p^{w_i}-X_p^{w_j}\right)^2\right),  \theta_p\in\mathbb{R}^+
    \label{eq:kernel}
\end{equation}
This kernel relies on $n'$ hyperparameters $\theta_p$ estimated by maximum of likelihood such that the more the number of variables $n'$ for the problem, the more the number of hyperparameters to optimize. 
Reducing this number leads to a better estimation for the hyperparameters, a more convenient optimization of the likelihood and makes the model scalable for high-dimensional problems. 
To do so, the Partial Least Squares (PLS) method~\cite{wold_1975} searches the direction that maximizes the variance between the input and output variables. This is done by a projection into a smaller space spanned by the so-called principal components. The number of principal components $d$ that corresponds to the new number of hyperparameters for KPLS is chosen to be much lower than $n'$. The resulting PLS squared exponential kernel is given by
\begin{equation}
k(X^{w^i},X^{w^j},\hat{\theta}) =  \prod\limits_{q=1}^{d} \prod\limits_{p=1}^{n'}\exp\left(-\hat{\theta}_q\left(b_{*p}^{q}X_p^{w_i}-b_{*p}^{q}X_p^{w_j}\right)^{2}\right),  \forall\ \theta_q\in\mathbb{R}^+
\label{eq:kernelPLS}
\end{equation}
where  $[b_{*p}^q]_{p,q}$ are scalars that measure the influence of the input variables on the output $y^f$.
Combining this model construction with SEGO gives the method described in Algorithm~\ref{alg:method}.

\smallskip
\begin{algorithm}[H]
\SetAlgoLined
\KwResult{Solution of the problem \eqref{eq:opt_prob} over the mixed categorical design space $\Omega \times S \times \mathbb{F}^l$.}
{\textbf{Inputs:}}  Initial DoE $\mathscr{D}_0$ and set $t=0$. The search space $ \Omega \times S \times \mathbb{F}^l $. 

\While{the stopping criterion is not satisfied}{
\vspace{.2cm}
\begin{adjustwidth}{0pt}{40pt}
\begin{enumerate}
     \item Relax continuously integer and categorical input variables to a real bounded space $\Omega'$ of dimension $n'= n+m+ \sum_{j=1}^l   L_j$. Namely, we continuously relax the mixed categorical DoE $\mathscr{D}_t=\{w^i\}_i \in (\Omega \times S \times \mathbb{F}^l)^{n_t} $ to a continuous DoE  $\mathscr{D}_t^{'}=\{X^{w^i}\}_i  \in (\Omega')^{n_t}$ using the relaxation procedure $Relax$.
    \item Build the GP models for the objective function $f$ and the constraints $g$ and $h$ related to the DoE $\mathscr{D}_t^{'}$ with PLS to reduce the number of the hyperparameters from $n'$ to $d_f,d_g$ and $d_h$ as detailed in~\algref{alg:woldR}.
    
    \item Build an estimation of the feasible domain $\Omega'_f \subset \Omega'$  with the criterion $UTB$ and construct the acquisition function $WB2s$.
    \item Maximize the acquisition function $WB2s$ over $\Omega_f'$: $X_{t}=  \underset{X \in \Omega_f'}{\arg \max}\ WB2s(X)$. 
    \item Project the obtained continuous solution over $ \Omega \times S \times \mathbb{F}^l $: $w_t=Project(X_t)$.  
    \item Add $w_{t}, f(w_{t}), g(w_{t}), h(w_{t})$ to the DoE  $\mathscr{D}_{t+1}$. Increment $t$.
\end{enumerate}
\end{adjustwidth}
\smallskip

} % End while

\caption{SEGO for mixed categorical inputs using an adaptive KPLS. %and Wold's R criterion.
}
\label{alg:method}
\end{algorithm}

\smallskip

%\paragraph{Choosing automatically the number of components}
KPLS is an efficient method that can tackle high-dimensional problems by reducing the number of effective dimensions to a small number and it is currently used in the community~\cite{bhosekar2018advances,zuhal2021dimensionality,li2021improved}. The number of principal components being a key point that must be chosen for the PLS technique, we propose a strategy to choose this number in an adaptive process through the optimization iterations. As a rule of the thumb, taking 2 to 5 active components is, in general, efficient for most problems. However, when we do not have any $\textit{a-priori}$ information, an intuitive idea is to learn from GP the number of active dimension. Bouhlel et al.~\cite{Bouhlel2016improving} proposed to use the leave-one-out strategy to find the best number of hyperparameters for KPLS.
However, this strategy is not efficient because at every iteration, it implies to  compute a large number of reduced size surrogate models to select the best one. As the size of the DoE increases, both the model computation and the leave-one-out criterion become more and more costly. 
In~\cite{PLS_Li_2002}, they propose to use the adjusted Wold’s R criterion~\cite{WoldPLS} for dimensionality reduction purposes. Wold’s R criterion is based on cross-validation over $k$-folds \cite{PLS_Li_2002} and consists in measuring  the ratio of the PRedicted Error Sum of Squares (PRESS)~\cite{PRESS}. If for a given number of components $d$, the criterion $R$ defined as 
\begin{equation}\label{eqPRESS}
R(d)=\frac{\text{PRESS}(d+1)}{\text{PRESS}(d)}
\end{equation}
and should be smaller than a given threshold. The ratio is typically 1, which means we add components until the approximate error stops decreasing. However a threshold of 0.9 or 0.95 could be used in order to be more selective and add fewer components. 

\smallskip
\begin{algorithm}[H]
\SetAlgoLined
\KwResult{Optimal model of reduced order}
{\textbf{Inputs:}} A DoE of size $n_t$, i.e, $\mathscr{D}= \{(x_1,y_1), \ldots, (x_{n_k},y_{n_t}) \} $ associated with a given function $\Psi$ ($\Psi(x_i)=y_i, \forall i=1,\ldots, n_t$). The maximal and minimal number of components $d_{\max}$ and $d_{\min}$, respectively. A threshold $\sigma  \in [0,1]$. A chosen number of folds $K$. $\forall k \in \{1, \ldots, K \}$ denote the corresponding known $n_k$ points: $\{(x_1,y_1), \ldots, (x_{n_k},y_{n_k}) \}$.

\textbf{Initialization:} 
\begin{itemize}
 \item For all subsamples $k \in \{1, \ldots, K \}$, build a KPLS model with $d_{\min} $ components over all data except the inputs in the subsample $k$. 
    \item For all $ i \in 1, \ldots, n_k$, let $\hat{y}_{i,-k}$ be the prediction of the KPLS model build without the point $i$ in the $k^{th}$ fold at the corresponding $x_i$. This fold is therefore used as a validation set to compute the $K$-fold cross validation PRESS as $ \text{PRESS}(d_{\min}) = {\displaystyle \sum_{k=1}^{K}{\displaystyle \sum_{i=1}^{n_k} \left(  y_i - \hat{y}_{i,-k}     \right)^2  }} $.
    \item Set $d_{\Psi} \xleftarrow{} d_{\min} $ .
\end{itemize}

\While{$d_{\Psi} \ < \  d_{\max}$}{
\vspace{.2cm}
\begin{adjustwidth}{0pt}{40pt}
\begin{enumerate}
    \item Divide  the DoE $\mathscr{D}$ into $K$ subsamples. 
  
    \item For all subsamples $k \in \{1, \ldots, K \}$, build a KPLS model with $d_{\Psi}$ components over all data except the inputs in the subsample $k$. 
    \item For all $ i \in 1, \ldots, n_k$, let $\hat{y}_{i,-k}$ be the prediction of the KPLS model build without the point $i$ in the $k^{th}$ fold at the corresponding $x_i$. This fold is therefore used as a validation set to compute the $K$-fold cross validation PRESS as $ \text{PRESS}(d_{\Psi}+1) = {\displaystyle \sum_{k=1}^{K}{\displaystyle \sum_{i=1}^{n_k} \left(  y_i - \hat{y}_{i,-k}     \right)^2  }} $. Set $R(d_\psi)=\frac{\text{PRESS}(d_{\Psi}+1)}{\text{PRESS}(d_{\Psi})}$.
    \item 
     \uIf{$R(d_\Psi) \geq \sigma $}{
   \quad  \textbf{STOP};
   }
   \item Increment $d_{\Psi}$, i.e., $d_{\Psi} \xleftarrow{} d_{\Psi}+1$  .
  
 \end{enumerate}

\end{adjustwidth}
\smallskip
} % End while
\textbf{Return} KPLS model with $ d_{\Psi}$ components;
   
\caption{%A framework for 
Adaptive dimension reduction for KPLS models.}
\label{alg:woldR}
\end{algorithm}

\smallskip
Note that when $d_{\min} =d_{\max}$, Algo.~\ref{alg:woldR} is equivalent to KPLS with a constant number of dimensions. The code implementation of our proposed method has been released in the toolbox SMT v1.1\footnote{\url{https://smt.readthedocs.io/en/latest/}}~\cite{SMT2019}.

%%%%%%%%%%%%%%%%%%%%%%%%%%%%%%%%%%%%%%%%%%%%%%%%%%%%%%%%%%%%%%%%%%%%%%

\section{Results and discussion}
\label{sec:res}
In this section, we carry out experiments for unconstrained and constrained test cases, with several number of variables and an increasing complexity.
We optimize analytical test cases as a benchmark study and then we solve aircraft design optimization problems as a validation and application of the present work.
\subsection{Implementation choices}

In order to compare, we used several optimization algorithms detailed hereafter: Bandit-BO, NSGA2, SEGO with Kriging, SEGO with Gower distance and the proposed method, SEGO coupling Kriging and PLS.

The Bandit-BO implementation used is the one by Nguyen et al.~\cite{Bandit-BO}, we are not considering parallelization or batch evaluations. 
The NSGA2~\cite{NSGA2} algorithm used is the implementation from the toolbox pymoo~\cite{pymoo} with the default parameters (probability of crossover = 1, eta = 3). Fronts are not relevant in our study as we are considering single-objective optimization. 
The optimization with SEGO is made from SEGOMOE~\cite{Bartoli2019} for both constrained and unconstrained  cases.
For SEGO using Gower distance~\cite{Gower} (denoted by SEGO+GD), we are considering the implementation of the Surrogate Modeling Toolbox (SMT)~\cite{SMT2019}, an open-source python toolbox where some computations associated to the present work have been done and where automatic PLS was implemented. 
The same holds for SEGO and Kriging coupled with PLS to reduce to $d$ the number of hyperparameters  (denoted by SEGO+KPLS(d=$d$). As the PLS could potentially lead to numerical instabilities, we are using the homoscedastic noise that maximizes the likelihood as a so-called nugget. For automatic PLS (denoted by SEGO+KPLS(auto), the $K$-fold cross-validation uses $K=4$, the experiments~\cite{PLS_Li_2002} suggest to use 4 to 6 fold cross-validation.

For SEGO using Kriging (denoted by SEGO+KRG), we also use the implementation from the toolbox SMT.  For the constrained analytical test cases, we are using the $UTB$ criterion~\cite{SEGO-UTB}.
Some adaptions have been done to Bandit-BO and NSGA2 to consider both integer and categorical variables. As NSGA2 can only consider integer variables, categorical variables are treated as integer ones. 
Contrarily, Bandit-BO can treat only categorical variables so integer variables are treated as categorical ones.
Bandit-BO creates a GP model for each arm, so it requires at least $2\times N_c$ initial points, $N_c$ being the number of categorical possibilities for the problem inputs. For unconstrained test cases, these $2$ points by categorical possibility are sampled randomly. 
If we are not using Bandit-BO, for constrained optimization, we are using a continuous relaxed Latin hypercube sampling and then we project the output points to obtain the mixed integer DoE.

For Kriging, Kriging with PLS or Gower distance, the hyperparameters are optimized with COBYLA~\cite{COBYLA} and the chosen model regression is constant. When optimizing with SEGO, the acquisition function is maximized using ISRES~\cite{ISRESimp} to find some interesting starting points and SNOPT~\cite{SNOPT7} to finalize the process based on these starting points.
The squared exponential kernel is the only one considered for these methods. 

In order to compute some statistical data (median and variance), we are doing 20 repetitions of the optimization process for a given method and an initial DoE size. We consider that a constraint is respected if the constraint violation is smaller than the threshold value $10^{-4}$.

\subsection{Benchmark test cases including unconstrained and constrained optimization problems}

\label{subsec:res_aca}
In this section, a benchmark of different problems is proposed in order to compare the efficiency of the proposed algorithm with some state-of-the-art methods. The first ten are analytical cases, with or without constraints in order to provide some data profiles and the last two concern some more complex applications for aircraft conceptual design. 
\subsubsection{Unconstrained optimization}
\begin{figure}[H]
   \begin{subfigure}[b]{.5\linewidth}
      \centering
	\includegraphics[clip=true,height=5cm]{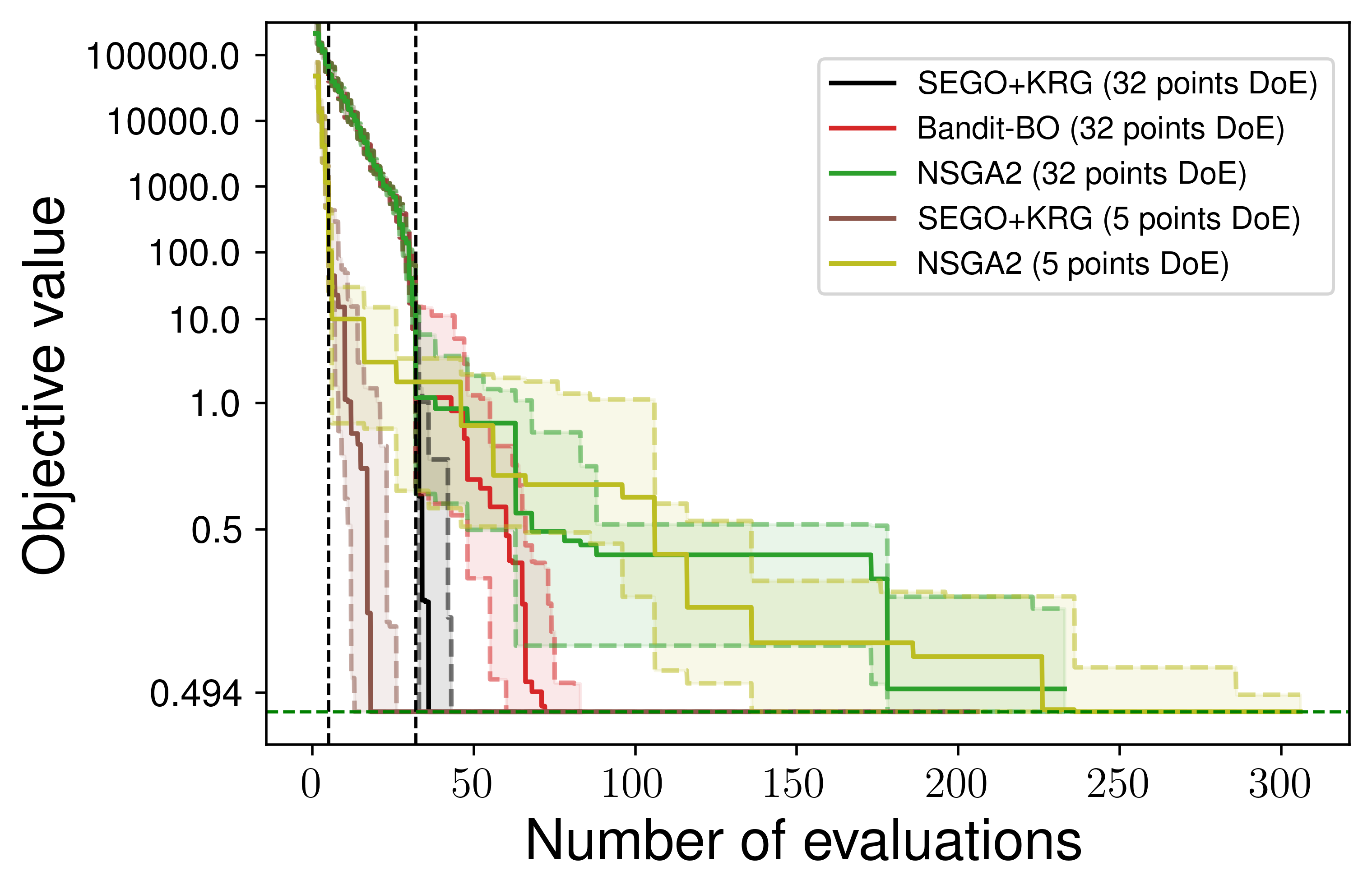}
     \caption{Convergence curves.}
     \label{B5testconv}
      \end{subfigure}
      \begin{subfigure}[b]{.5\linewidth}
      \centering 
	\includegraphics[clip=true,height=5cm]{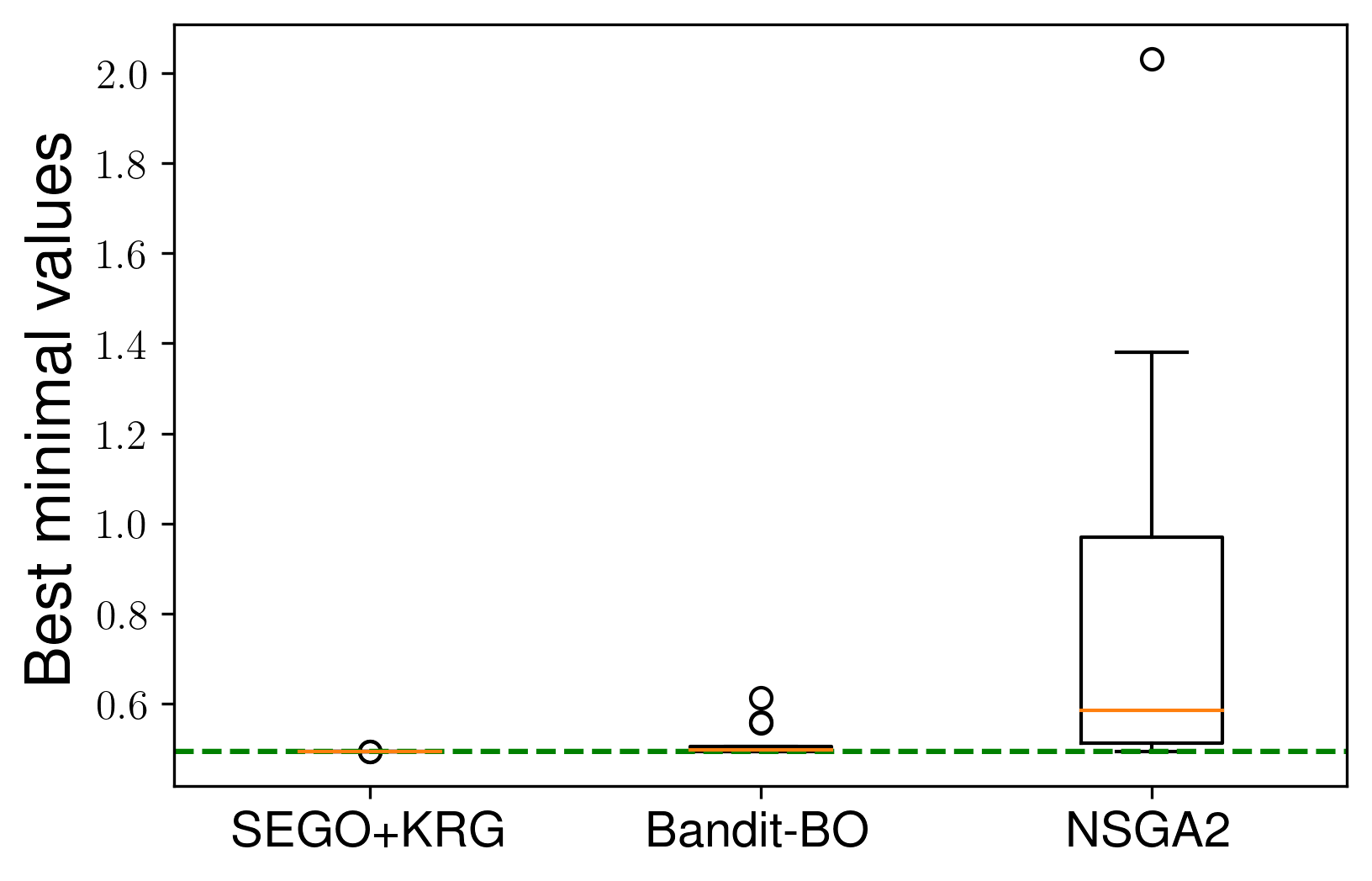}
     \caption{Boxplot with a DoE of 32 points.}
     \label{B5testboxplots50}
   \end{subfigure}
   \caption{``\texttt{Branin 5}'' obtained optimization results. The Boxplots are generated, after 50 iterations, using the 20 best points. }
   \label{S1optim}
\end{figure}

To begin with, we validate our method on unconstrained problems up to 14 dimensions.
The first analytical test case is a modified Branin function~\cite{AMIEGO19}, denoted by ``\texttt{Branin 5}'', where the first variable is an integer  $x_1 \in \{ -5,-4,\dots, 9, 10 \}$ and the second one $x_2 \in [0,10]$ is a continuous variable.  As this problem is only 2-dimensional, SEGO-KRG is considered without the coupling with the PLS technique.
For Bandit-BO, we represent $x_1$ as a bandit with 16 arms associated to the 16 integer values from $0$ to $15$. Therefore, an initial DoE of at least 32 points is required for Bandit-BO. A smaller initial DoE is also considered for NSGA2 and SEGO-KRG in order to compare the convergence according to the DoE size: 5 points or 32 points. For a given DoE size (5 or 32), 20 different initial DoE are obtained via Latin hypercube sampling in order to compute some statistical data (median and variance) about the convergence results. 
In~\figref{B5testconv}, the medians and the associated quartiles (25\% and 75\%)  of the 20 runs are illustrated for Bandit-BO, NSGA2 and SEGO with Kriging. The initial DoE is shown before the  black dotted line.
For the DoE with 32 initial points, we are doing 50 iterations of the methods Bandit-BO and SEGO-KRG, for a total of 82 evaluations. For NSGA2, 200 iterations are performed.  When the DoE  size is reduced to 5 points, 200 iterations of SEGO-KRG are made and compared to 300 evaluations for NSGA2.  These comparisons show that SEGO-KRG outperforms BANDIT-BO and NSGA2 in terms of convergence. Moreover, we find that a smaller initial DoE performs better for SEGO, that is a known result~\cite{riche2021revisiting}.

%\begin{figure}[H]
%\centering
%	\centering	
%	\includegraphics[clip=true,width=0.75\textwidth]{figures/B5_32_50.png}
%     \caption{Optimization results convergence curves}
%    \end{figure} 
%
%\begin{figure}[H]
%\centering
%	\centering	
%	\includegraphics[clip=true,width=0.75\textwidth]{figures/B5_5_200.png}
%     \caption{Optimization results convergence curves}
%    \end{figure} 

To compare the dispersion of the best results over the 20 runs,  \figref{B5testboxplots50} shows the boxplots for SEGO-KRG, Bandit-BO and NSGA2 after 50 iterations associated to a initial DoE of 32 points. These plots show that, not only SEGO-KRG converges better in median but also it is the only method to converge for every of the 20 initial DoE. However, Bandit-BO converges better than NSGA2 on this test case.

The second analytical test case is a toy function that consists in the choice between a set of 10 1-dimensional continuous functions~\cite{CAT-EGO}, denoted by ``\texttt{Set 1}'', where the first variable is a categorical variable  $x_1 \in \{ 0,1, \dots,9 \}$  and the second one $x_2 \in [0,1]$ is a continuous variable.  
This problem only has 2 variables but, as the relaxed space is in dimension 11, SEGO with KPLS is then considered to compare with SEGO with Kriging such that we can see how reducing the number of spatial correlations from 11 to 5 in the model can impact the optimization process. 
For Bandit-BO, we represent $x_1$ as a bandit with 10 arms from $0$ to $9$ and an initial DoE with a minimum of 20 points is required.
As previously for NSGA2 and SEGO-KRG, two initial DoE sizes are considered with 5 points and 20 points. For a given DoE size, 20 different initial DoE are obtained to compute statistics on the convergence results.

\begin{figure}[H]
   \begin{subfigure}[b]{.5\linewidth}
      \centering
      \includegraphics[height=5cm]{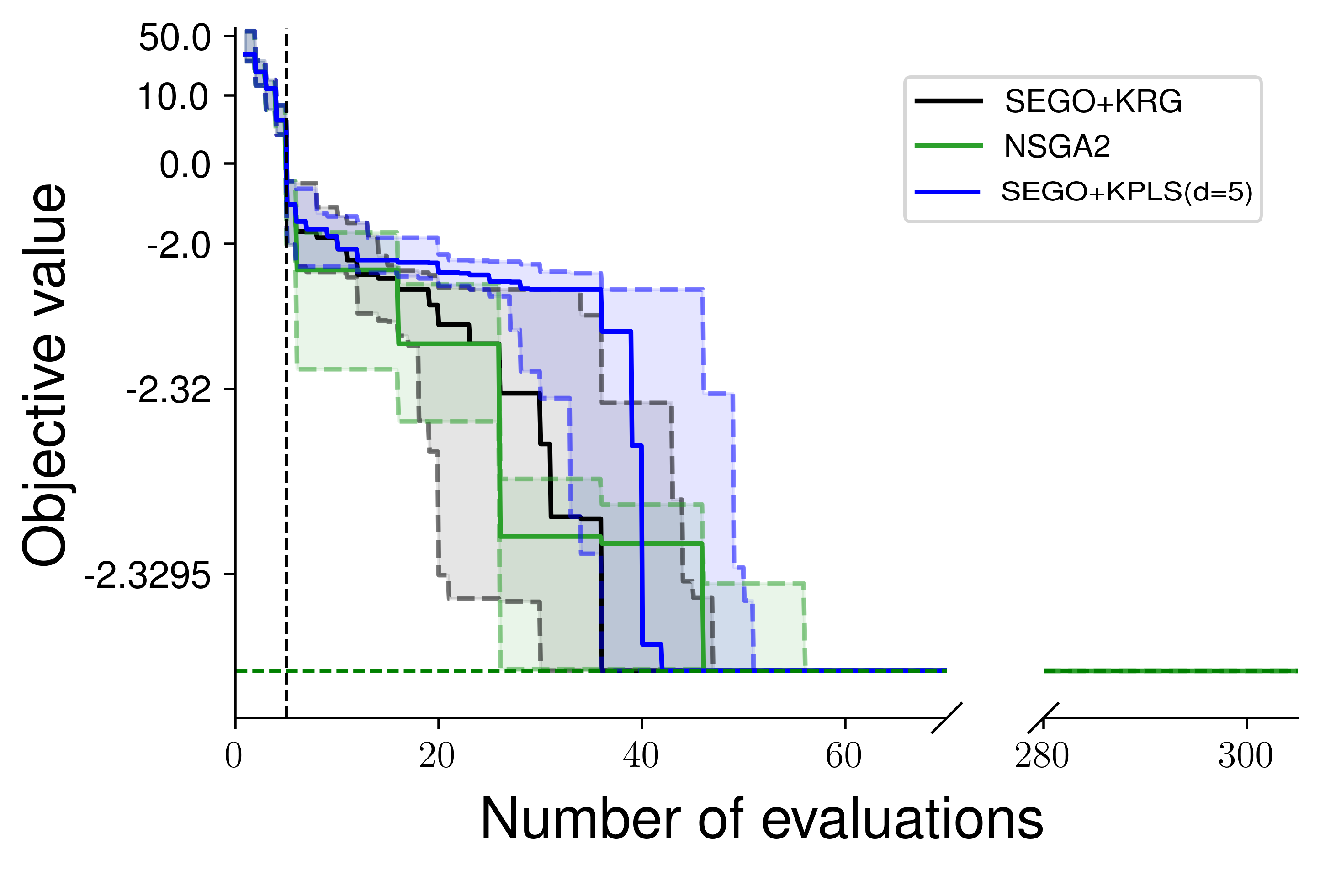}
     \caption{Convergence curves using 5 point initial DoE}
      \label{S1_5_200}
      \end{subfigure}
      \begin{subfigure}[b]{.5\linewidth}
      \centering 
      \includegraphics[height=5cm]{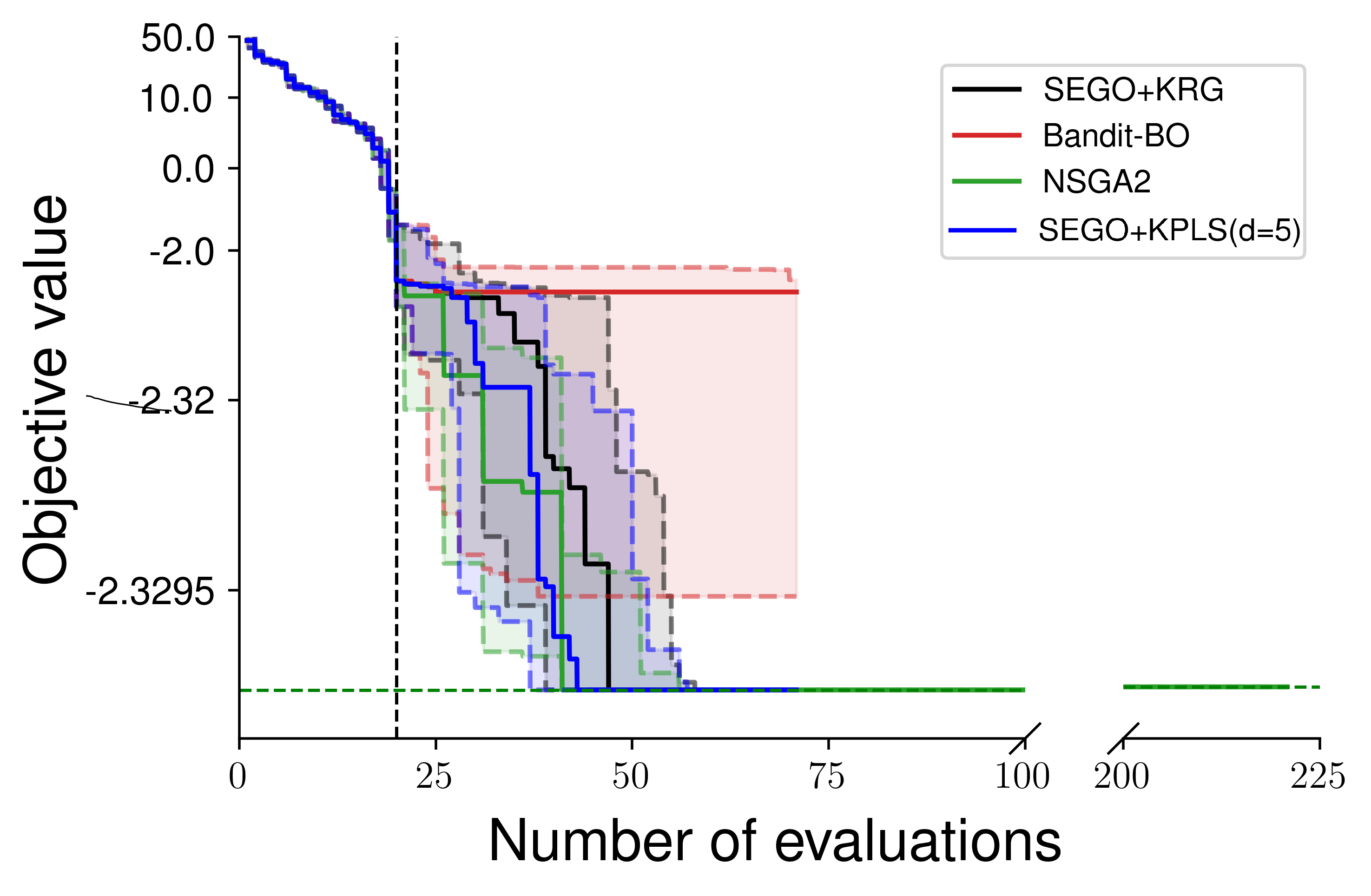}
      \caption{Convergence curves using 20 point initial DoE.}
      \label{S1_20_50}
   \end{subfigure}
    \begin{center}
    \begin{subfigure}{.5\linewidth}
      \centering 
 	\includegraphics[height=4cm]{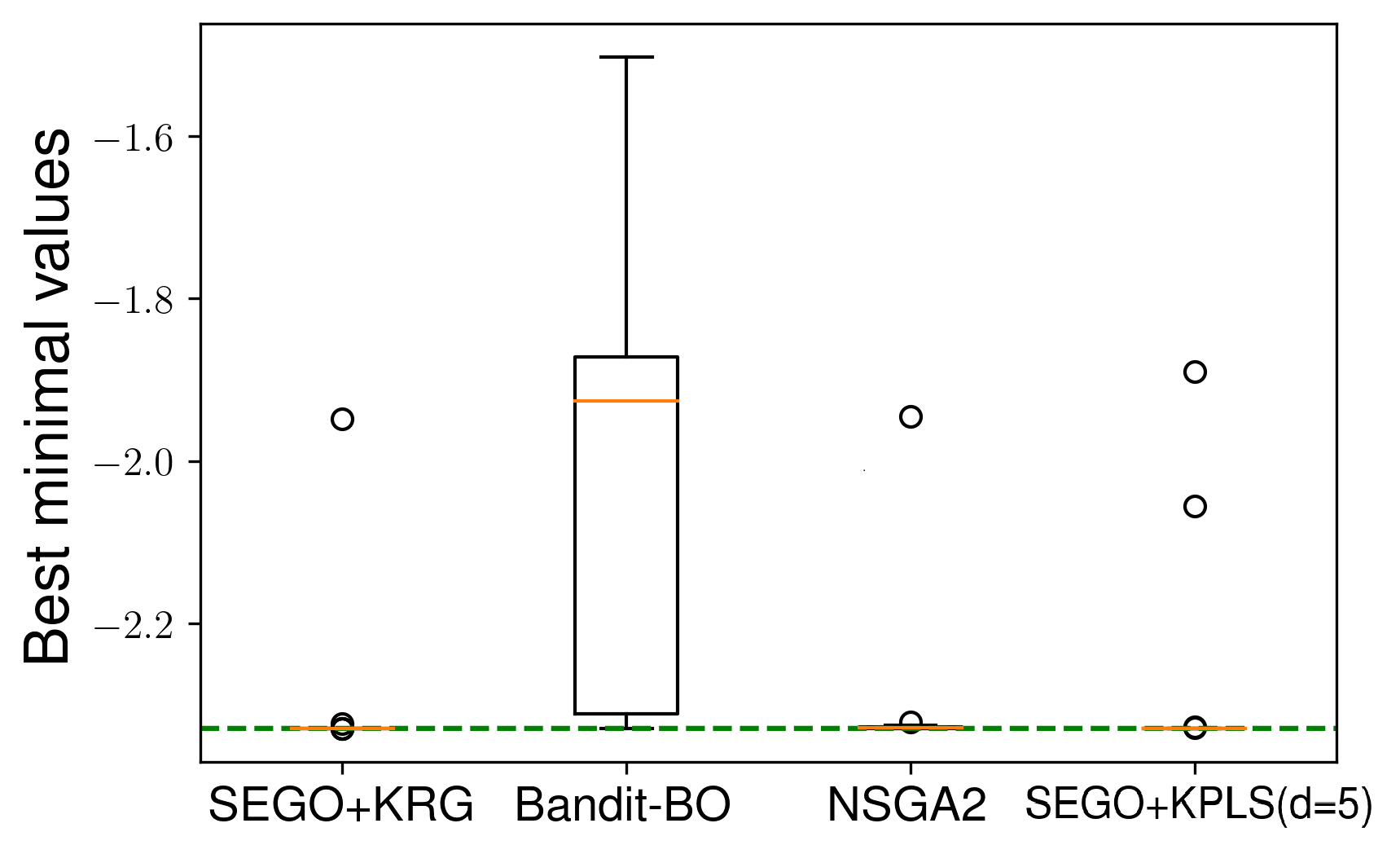}
     \caption{Boxplots with a DoE of 20 points.}
     \label{S1testboxplots50}
   \end{subfigure}
   \end{center}
   \caption{``\texttt{Set 1}'' obtained optimization results.}
   \label{S1optim}
\end{figure}

In~\figref{S1optim}, the medians and the associated quartiles (25\% and 75\%)  of the 20 runs are illustrated for each of the four algorithms. The initial DoE is shown before the black dotted line.
For the DoE with 20 initial points in~\figref{S1_20_50}, 50 iterations of the methods Bandit-BO, SEGO+KPLS(d=5) and, SEGO-KRG are performed for a total of 70 evaluations and for NSGA2 200 iterations are done.
When the DoE  size is reduced to 5 points in ~\figref{S1_5_200},  200 iterations of SEGO-KRG are done and compared to 300 evaluations for NSGA2.  These convergence plots show that the smaller the DoE, the faster the convergence. Also, KPLS slows the convergence at the start but this dimension reduction does not change the convergence overall and the incumbent is even better at the end with KPLS than without. On this test case, Bandit-BO method does not perform well.
To compare the dispersion of the best results over the 20 runs,  \figref{S1testboxplots50} shows the boxplots for SEGO-KRG, Bandit-BO, NSGA2 and, SEGO+KPLS(d=5) after 50 iterations associated to a initial DoE of 20 points. These plots show that SEGO+KPLS does not converge properly  because 2 runs are outliers that did not have converged whereas with NSGA2 and SEGO-KRG, there is only one outlier. Nevertheless, NSGA2 is not as precise as SEGO-KRG on average.

\subsubsection{Constrained optimization}

\begin{figure}[H]
   \begin{subfigure}[b]{.5\linewidth}
      \centering
	\includegraphics[clip=true,,height=4.5cm]{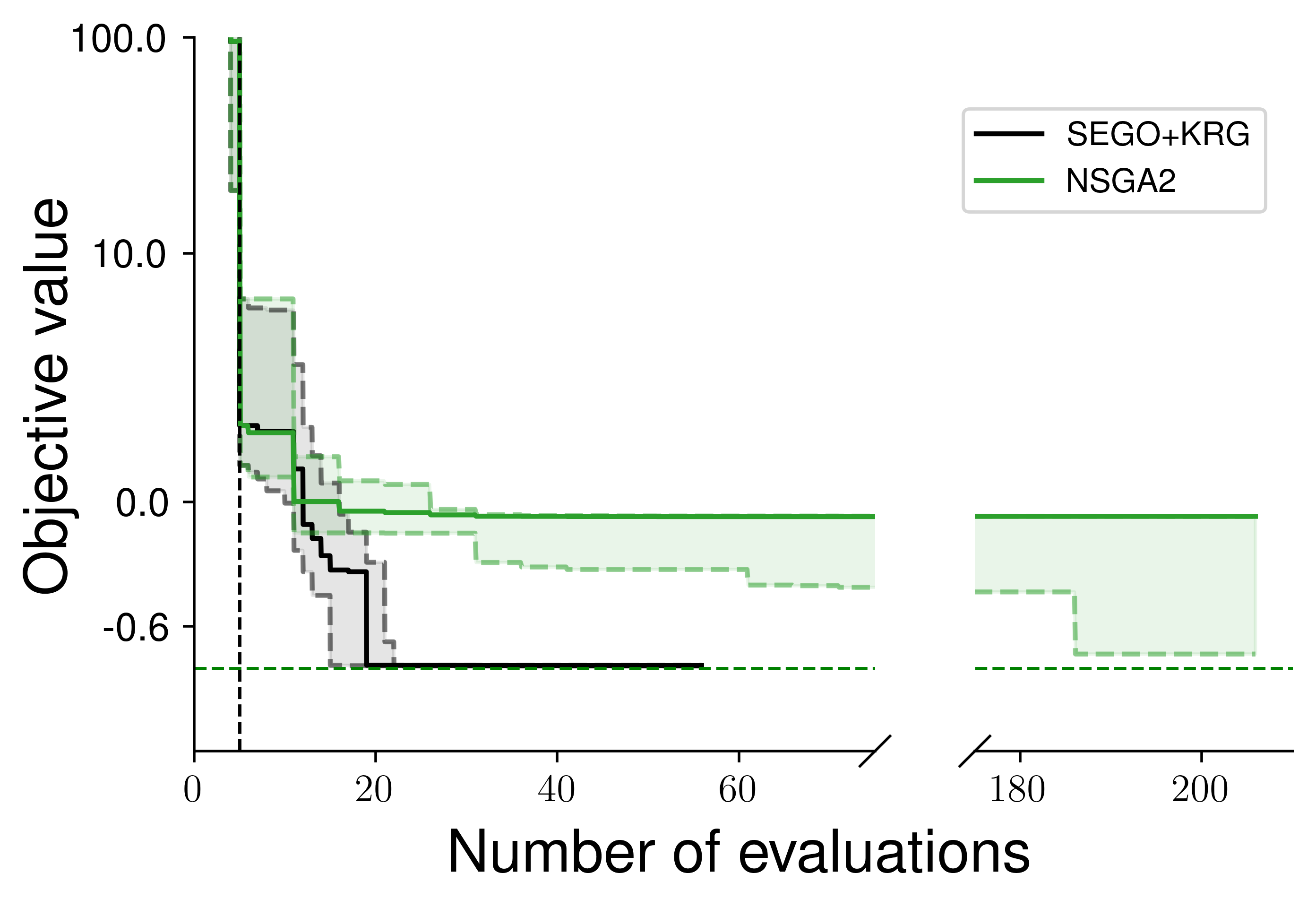}
     \caption{Convergence curves.}
     \label{B3testconv}
      \end{subfigure}
      \begin{subfigure}[b]{.5\linewidth}
      \centering 
	\includegraphics[clip=true,height=4.5cm]{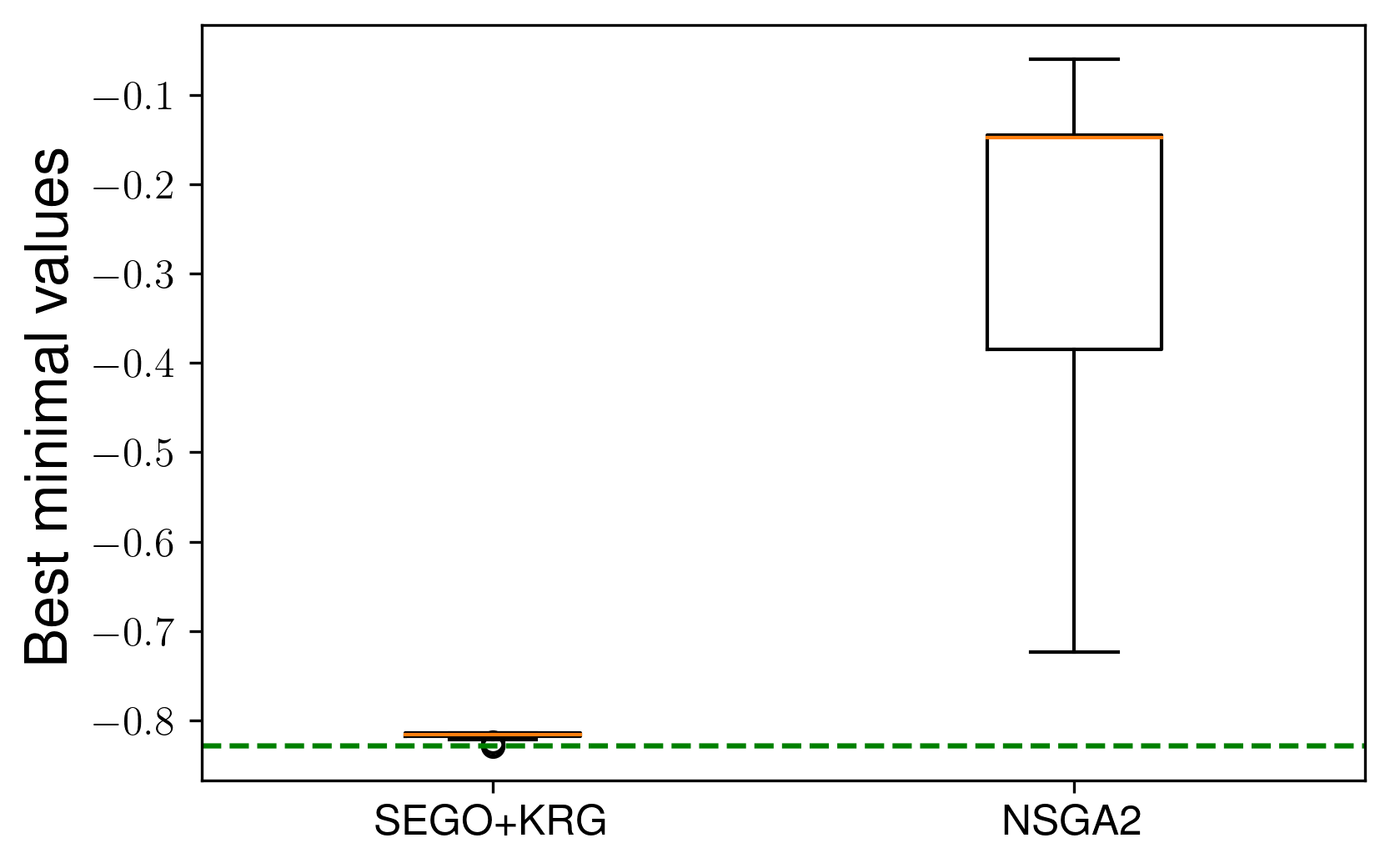}
     \caption{Boxplots.}
     \label{B3testboxplots50}
   \end{subfigure}
   \caption{``\texttt{Branin 3}'' optimization results.}
   \label{B3}
\end{figure}

For constrained optimization, Bandit-BO can not be considered as it can not deal with constraints, so SEGO is only compared to NSGA2.
The first constrained test case is a modified Branin function~\cite{Pelamatti} with one constraint, denoted by ``\texttt{Branin 3}'', where the two first variables are categorical variables with 2 levels each such that there is 4 possible Branin function variations and the two last variables are the continuous ones.
This problem  has 4 variables in the initial space and 6 in the relaxed one, so SEGO with Kriging is applied  without any dimension reduction technique.
According to the previous experiments, a small initial DoE with 5 points is considered in order to obtain better results for a given number of evaluations. 
In~\figref{B3testconv}, the medians and the associated quartiles (25\% and 75\%)  of the 20 runs are illustrated for NSGA2 and SEGO-KRG. The initial DoE is shown before the  black dotted line, then 50 iterations of SEGO-KRG are performed and compared with 200 iterations of NSGA2.
To compare the dispersion of the best results over the 20 runs,  \figref{B3testboxplots50} shows the boxplots for SEGO-KRG and NSGA2 after 50 iterations.
On this low-dimension constrained case, the mixed integer version of SEGO with Kriging is shown to perform well and be adapted to the constrained mixed optimization.

\begin{figure}[H]
   \begin{subfigure}[b]{.5\linewidth}
      \centering
	\includegraphics[clip=true,,height=5cm]{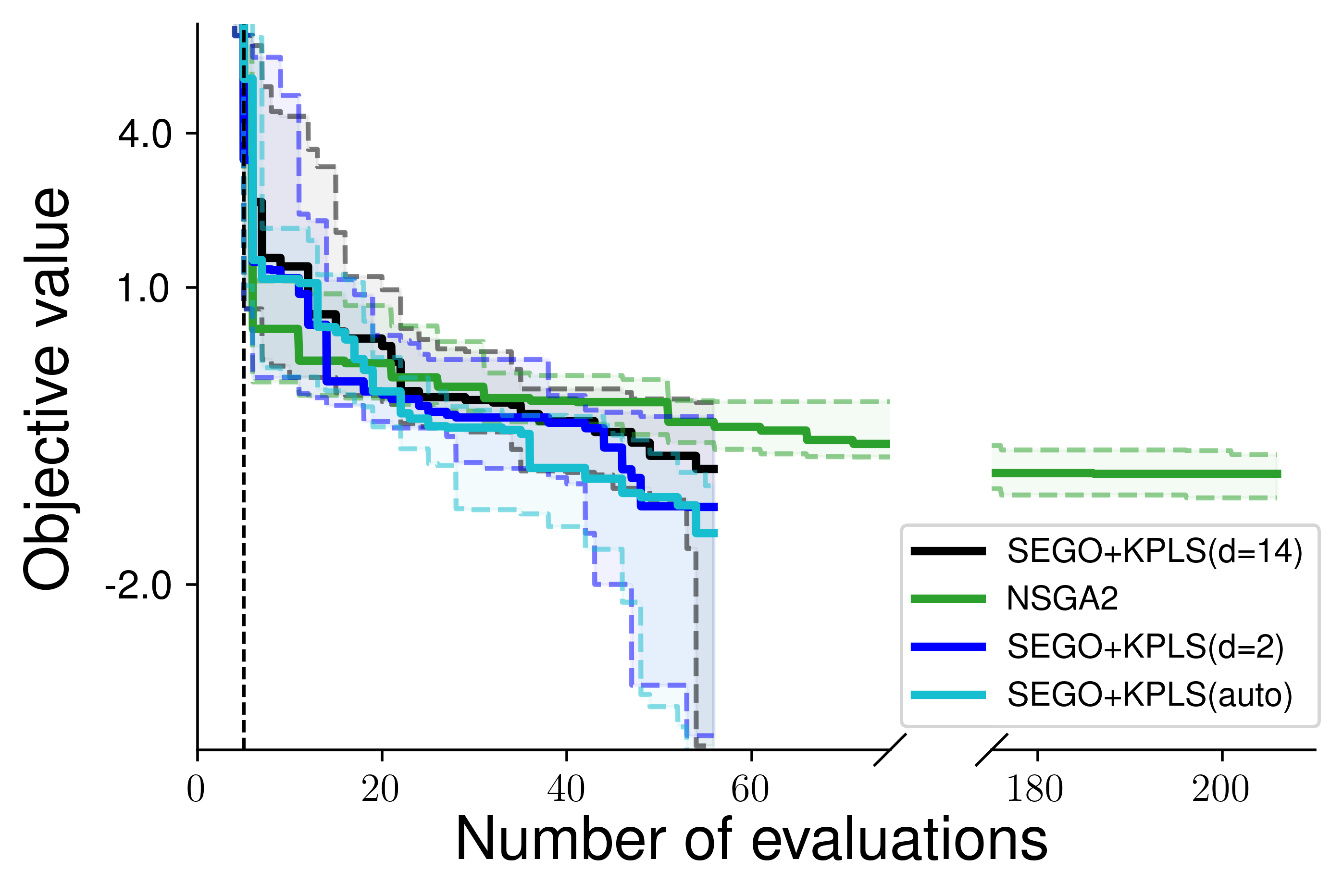}
     \caption{Convergence curves.}
     \label{B4testconv}
      \end{subfigure}
      \begin{subfigure}[b]{.5\linewidth}
      \centering 
	\includegraphics[clip=true,height=5cm]{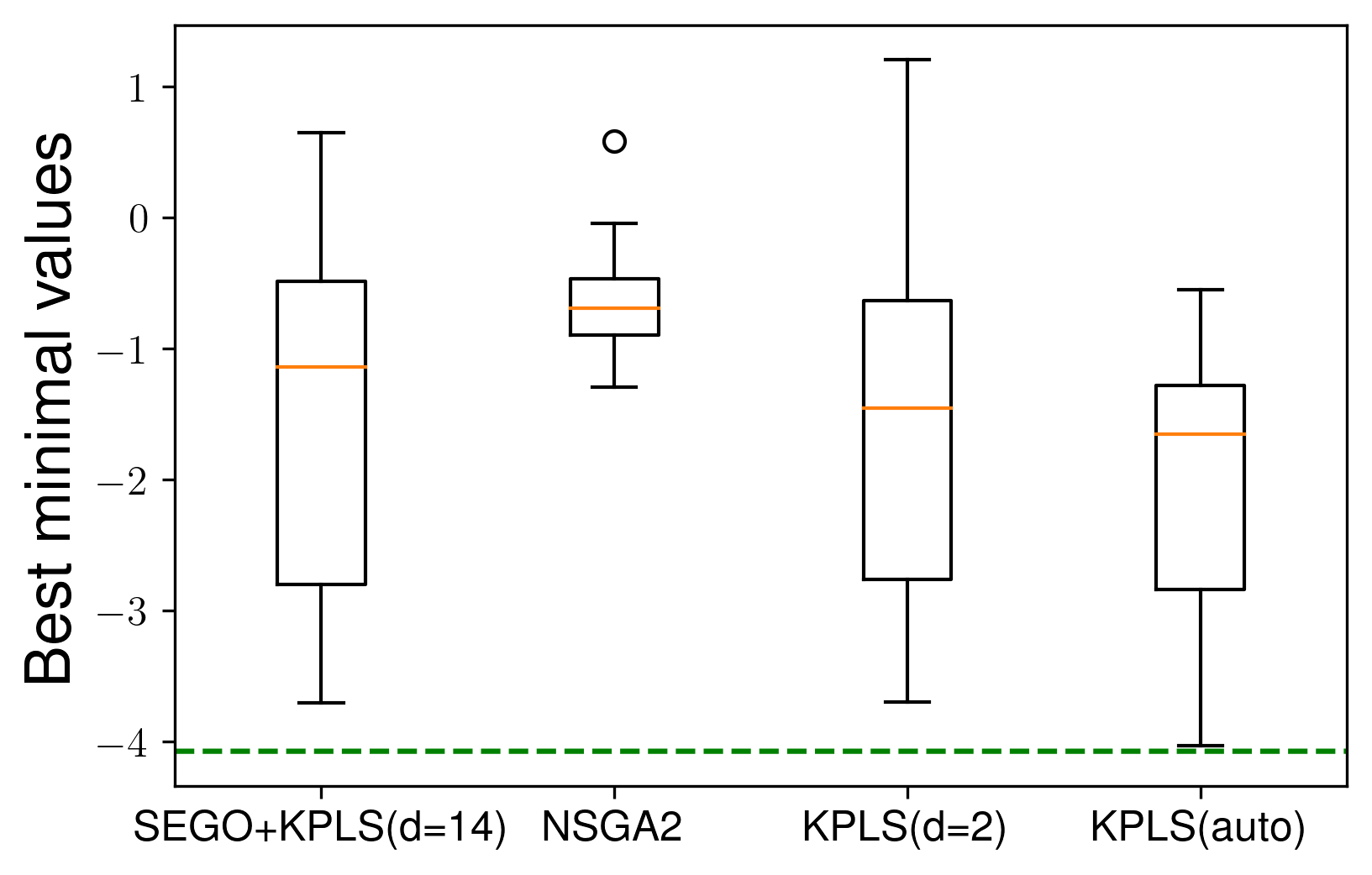}
     \caption{Boxplots.}
     \label{B4testboxplots50}
   \end{subfigure}
   \caption{``\texttt{Branin 4}'' optimization results.}
   \label{B4}
\end{figure}

\begin{figure}[H]
   \begin{subfigure}[b]{.5\linewidth}
      \centering
	\includegraphics[clip=true,,height=5cm]{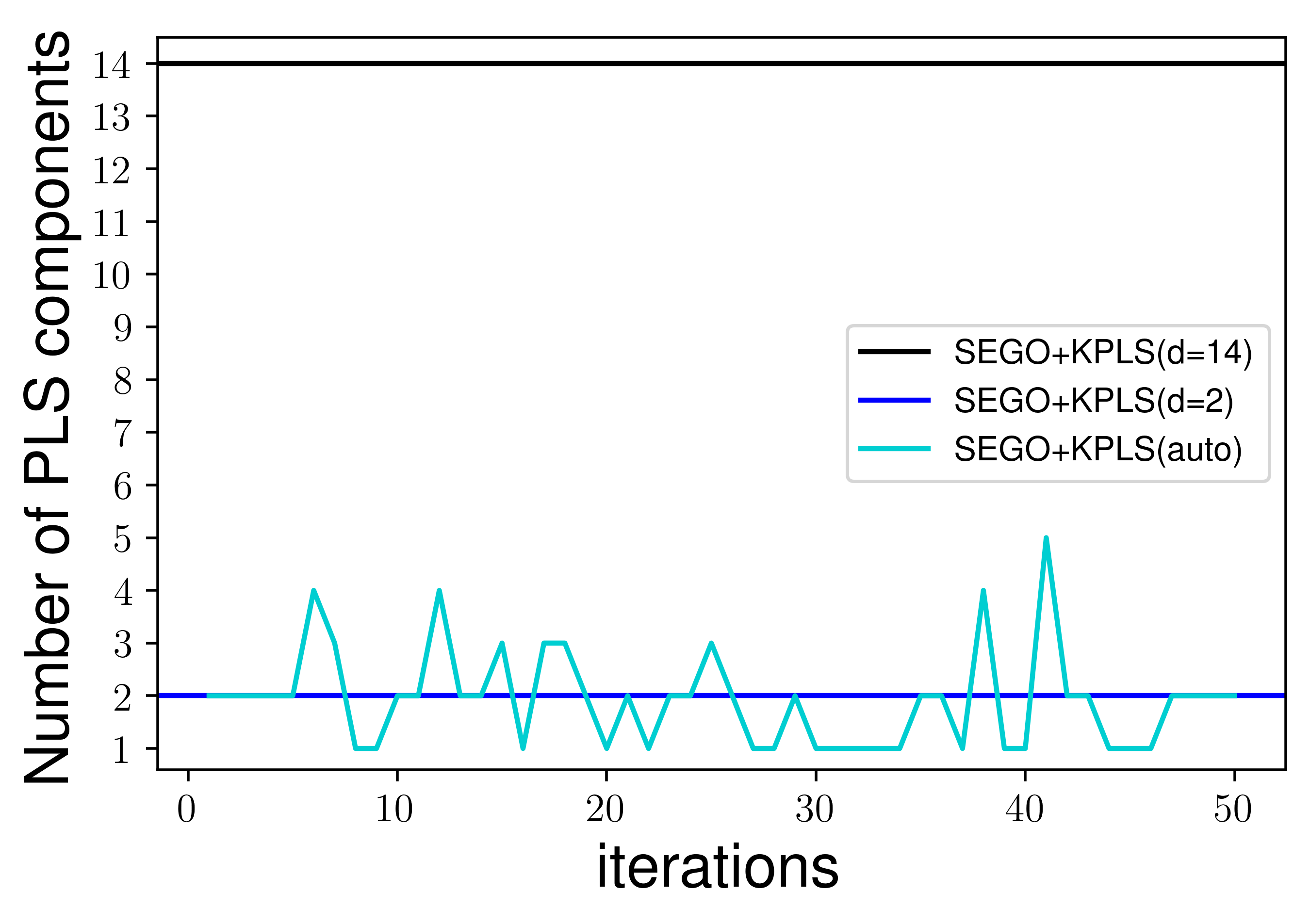}
     \caption{Model of the objective.}
     \label{B4modPLS}
      \end{subfigure}
      \begin{subfigure}[b]{.5\linewidth}
      \centering 
	\includegraphics[clip=true,height=5cm]{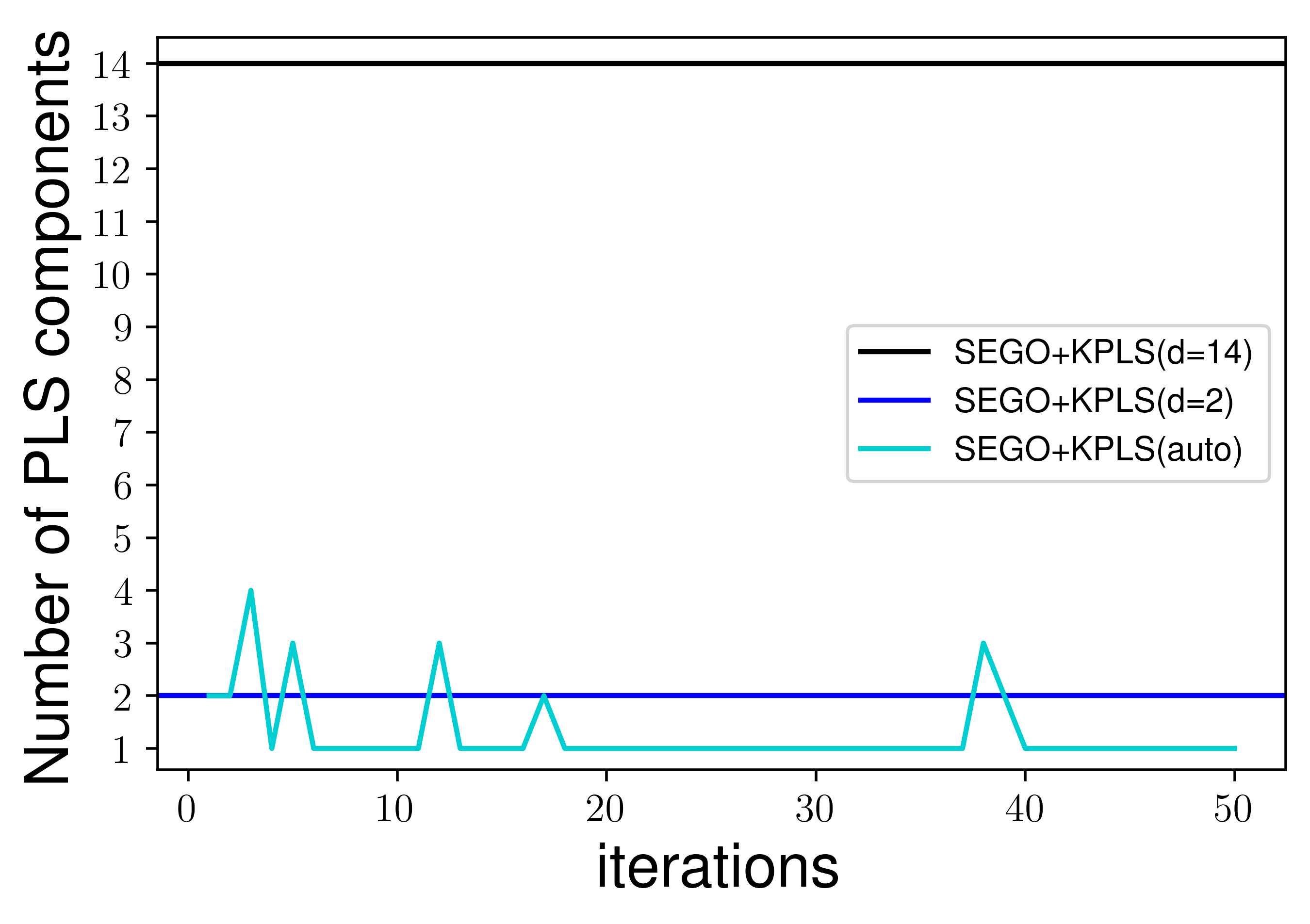}
     \caption{Model of the constraint.}
     \label{B4conPLS}
   \end{subfigure}
   \caption{``\texttt{Branin 4}'' PLS number of components.}
   \label{B4PLS}
\end{figure}

The second constrained test case is an augmented dimension modified Branin function~\cite{Pelamatti} with one constraint, denoted by ``\texttt{Branin 4}''. The first two variables are categorical with 2 levels each and the 10 last variables are continuous. Therefore, we can see the test case as a set of 4 continuous augmented 10-dimensional Branin functions. This problem  has 12 variables in the initial space and 14 in the relaxed space. SEGO with KPLS is considered with an automatic number of principal components to compare the dimension reduction with SEGO+KRG (equivalent to SEGO+KPLS(d=14)) and with SEGO+KPLS(d=2).
As previously, an initial DoE of 5 points is chosen, shown before the dotted line in~\figref{B4testconv}.  The medians and the associated quartiles (25\% and 75\%)  of the 20 runs are illustrated for NSGA2 and SEGO. 
On this graph, 50 iterations of SEGO+KRG are compared with 200 iterations of NSGA2.
To illustrate the dispersion of the best results over the 20 runs,  \figref{B4testboxplots50} shows the boxplots for SEGO+KPLS(d=14), SEGO+KPLS(d=2), SEGO+KPLS(auto) and NSGA2 after 50 iterations.
On this high-dimension constrained case, our method is still efficient. SEGO+KPLS(auto) and SEGO+KPLS(d=2) give similar results. To see how the different PLS models behave on the first run, we plotted on~\figref{B4PLS} the number of components through the 50 iterations. The adaptive method gives a mean number of 1.9 components for the objective model and of 1.25 components for the constraint model: the dimension reduction is quite efficient compared to KPLS(d=2) even in a few iterations  more components are necessary (up to 5 for the objective function). For both models, on~\figref{B4modPLS}, the estimated PRESS error reduction (based on~\eqnref{eqPRESS}) is reported through the iterations in which PLS components are added.  We can see that this error reduction increases with the size of the DoE. Due to the fact that a small DoE leads to a lot of uncertainties, adding components gives marginal gains with sparse data. Contrarily, when the number of known points is larger, adding components to the model increases the quality of its prediction (up to 10\%). These preliminary results on the `\texttt{Branin 4}'' test case motivate the interest for the adaptive method. 

\begin{figure}[H]
   \begin{subfigure}[b]{.5\linewidth}
      \centering
       \hspace*{-0.25cm}
	\includegraphics[clip=true,,height=5cm]{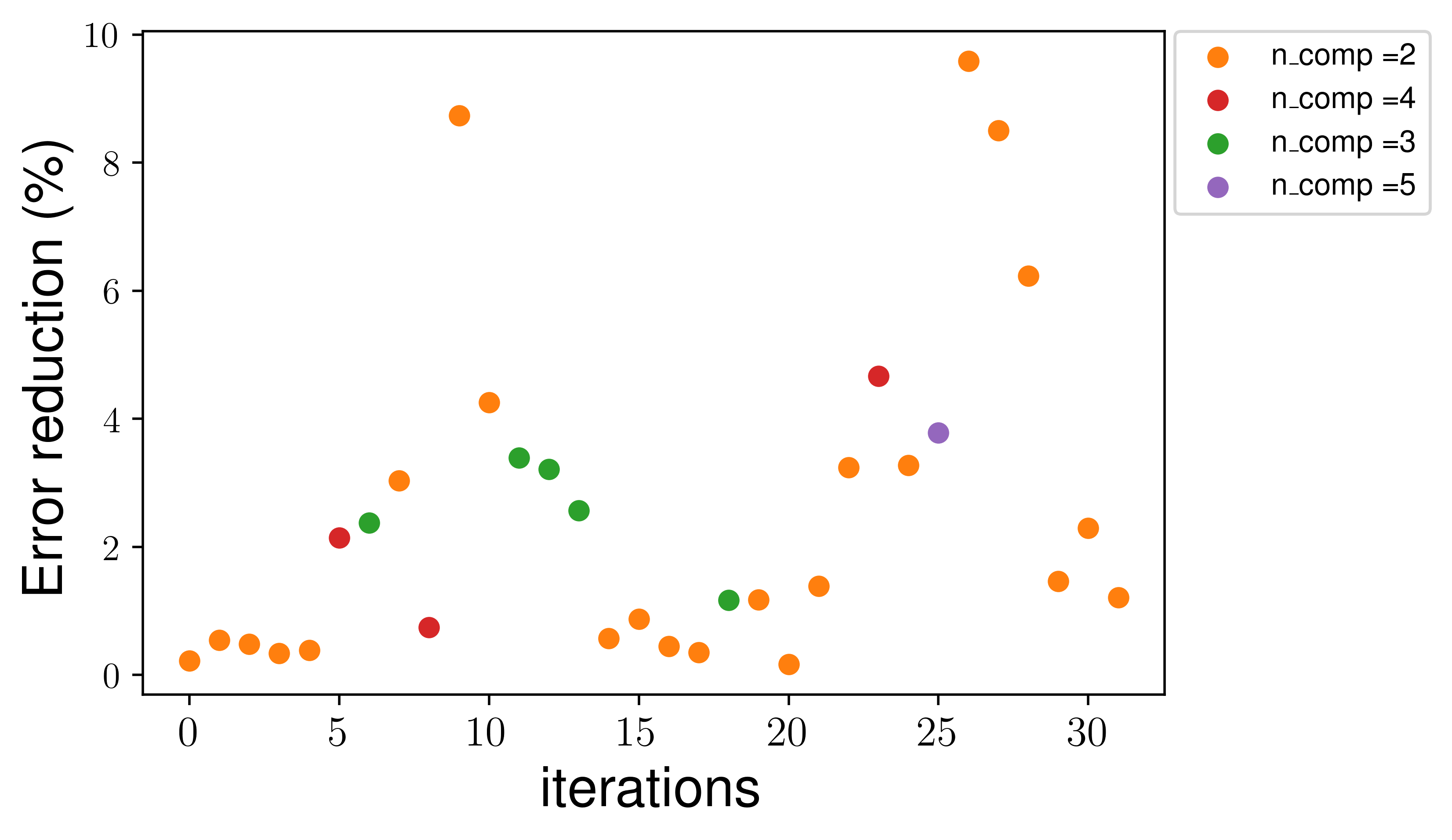}
     \caption{Model of the objective.}
     \label{B4modPLS}
      \end{subfigure}
      \begin{subfigure}[b]{.5\linewidth}
      \centering 
    \hspace*{0.25cm}\	\includegraphics[clip=true,height=5cm]{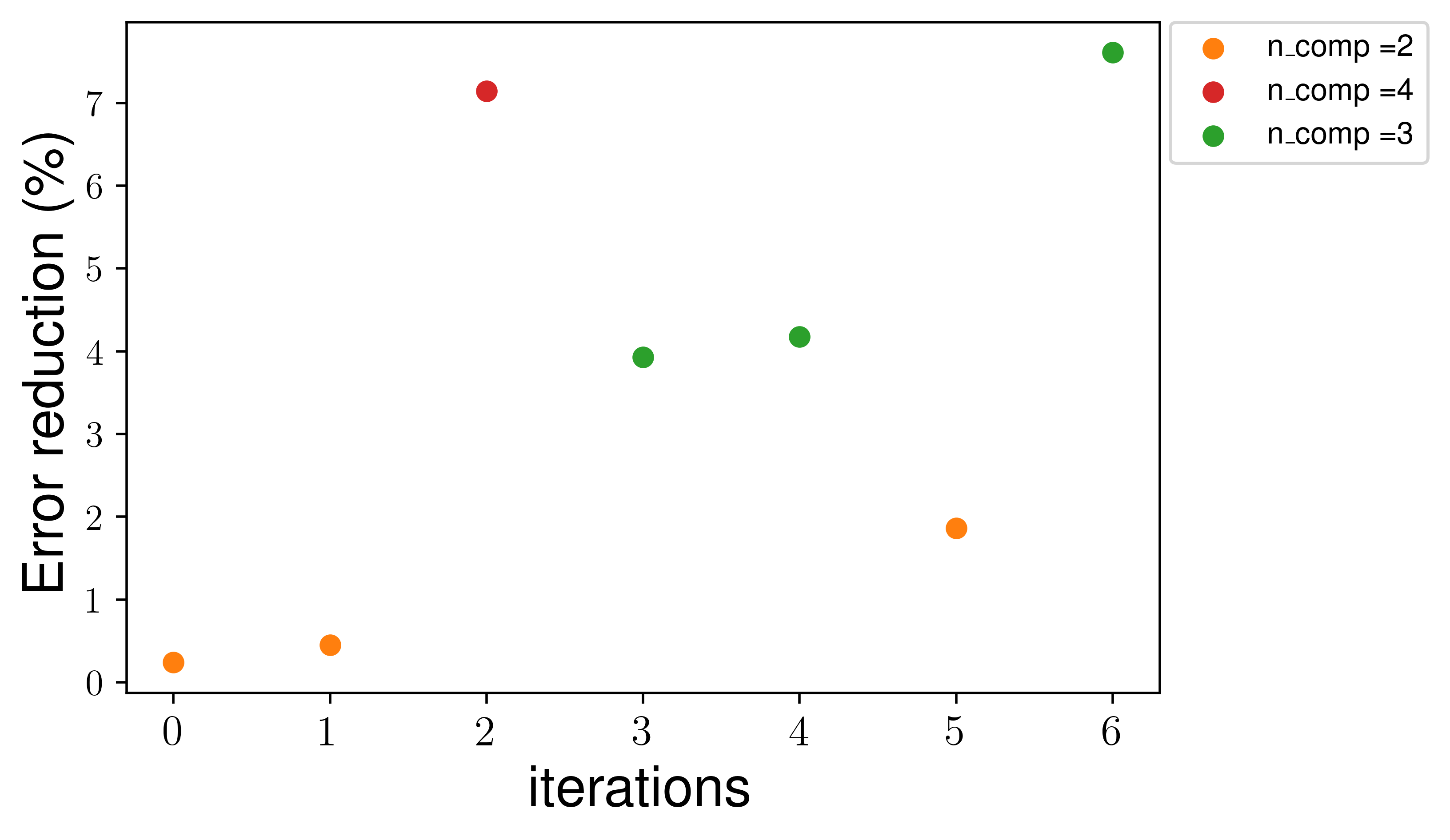}
     \caption{Model of the constraint.}
     \label{B4conPLS}
   \end{subfigure}
   \caption{Error gained when choosing more than 1 component.}
     \label{B4modPLS}
\end{figure}    

These four analytical cases have shown that SEGO performs better than both Bandit-BO for mixed integer Bayesian Optimization and NSGA2 for constrained optimization. It has been shown that the PLS technique allows the method to be scalable for high-dimension and even to give better results than Kriging by carrying favor to the most relevant space.

\subsubsection{Overall benchmark results} 
\begin{figure}[H]
\vspace*{-0.2cm}

   \begin{subfigure}[b]{.5\linewidth}
      \centering
	\includegraphics[clip=true,height=5cm]{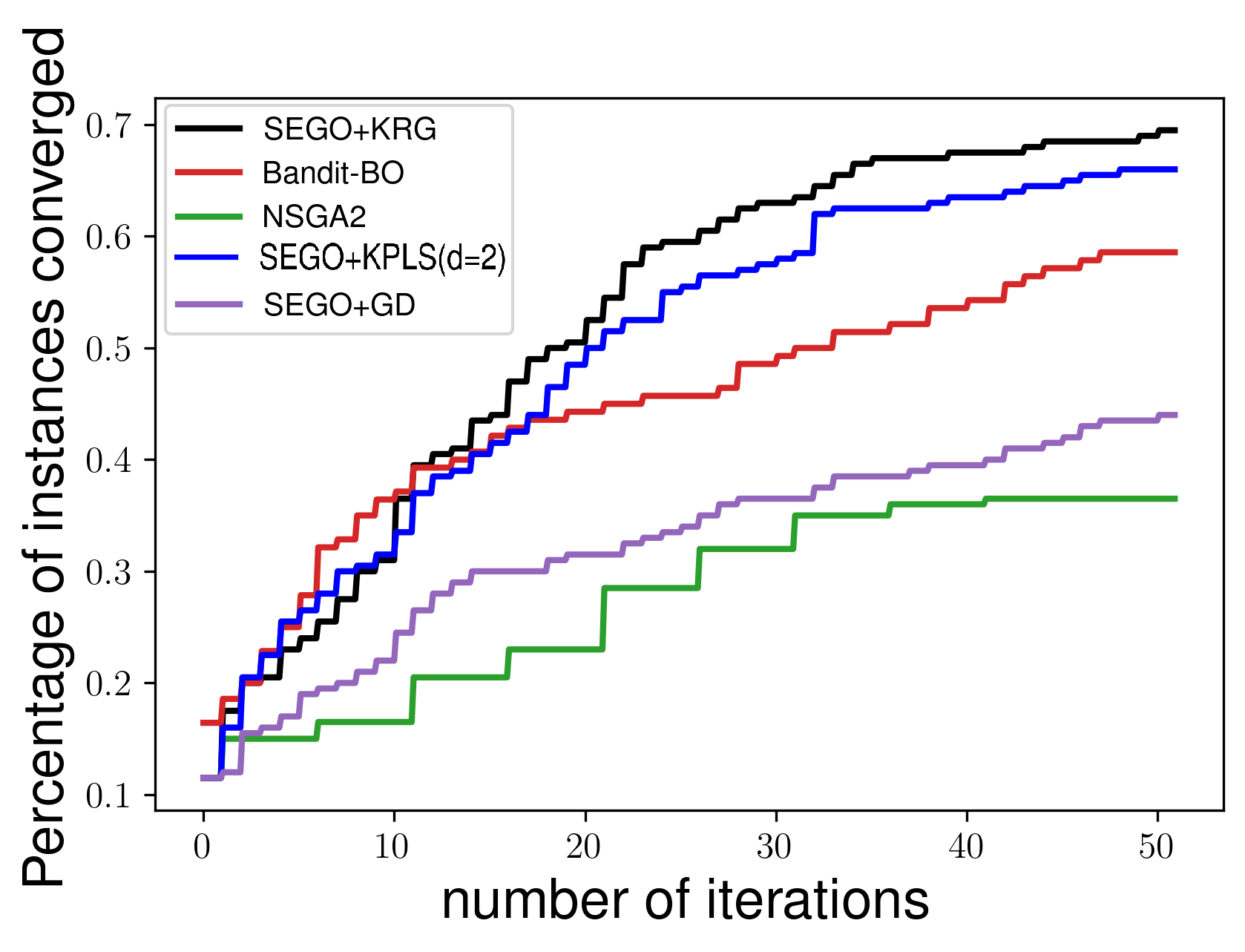}
     \caption{Data profiles for a tolerance of 2\%}
     \label{2DP}
      \end{subfigure}
        \begin{subfigure}[b]{.5\linewidth}
        \vspace*{-0.5cm}

      \centering
	\includegraphics[clip=true,height=5cm]{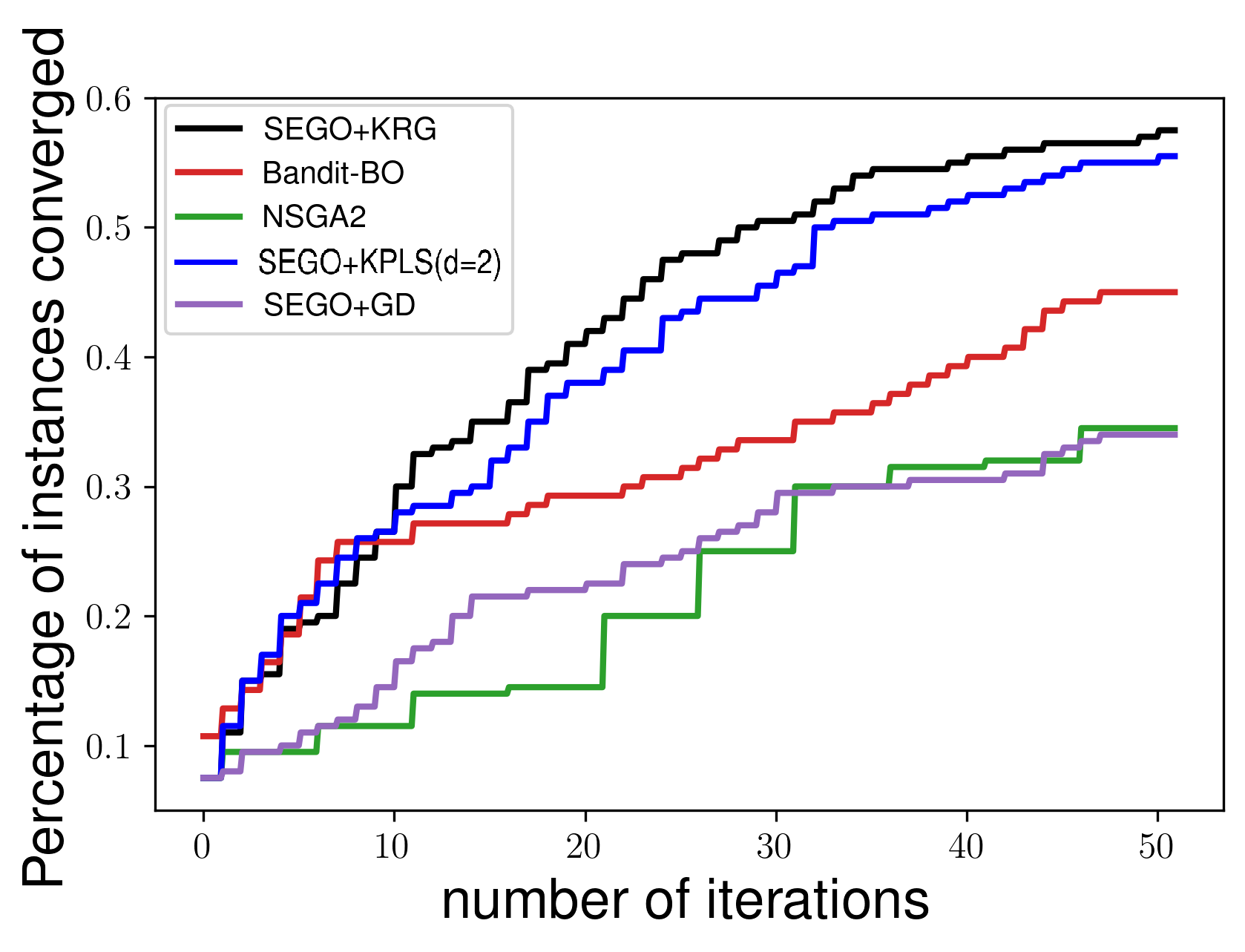}
     \caption{Data profiles for a tolerance of 0.5\%}
     \label{0.5DP}
      \end{subfigure} 
   \caption{Data profiles generated using 10 analytical test cases.}
     \label{DP_benchmark}
\end{figure}

For 10 different test cases (3 constrained and 7 unconstrained) we are considering 20 runs with different DoE sampled by Latin hypercube sampling for a total of 200 instances. 
 The resulting percentages of instances that have converged after a given budget for every method are plotted on~\figref{DP_benchmark}. For unconstrained test cases, in order to compare with Bandit-BO, the size of the initial DoE is given by $\min \{5, 2\times N_c \}$ where $N_c$ is the number of categorical possibilities and for constrained test cases, we took 5 points for the initial DoE size. This allow us to compare Bandit-BO, SEGO, SEGO with Gower Distance, SEGO with KPLS and NSGA2. Some tests are in dimension 2, so in order to compare with the same number of hyperparameters for all tests, we had to choose between 1 or 2 principal components for KPLS. As the number of points increases, the projected points are really closed to each other, so, to insure a certain stability, we choose KPLS with 2 principal components and noise evaluation, denoted by KPLS(d=2) in the following.

These 10 test cases were extracted from several state-of-the-art papers~\cite{Pelamatti,Pelamatti2020,Roustant18,CAT-EGO,AMIEGO19,Gower,Vanaret} from dimension 2 to 12, with integer, continuous and/or categorical variables. For constrained case, we keep only the inputs that give a constraint violation smaller than  $10^{-4}$.
A problem is considered solved if the error to the known solution is smaller than 2\% on~\figref{2DP} and smaller than 0.5\% on~\figref{0.5DP}. The mean error after 50 iterations can be found in~\tabref{tab:meanerror}. SEGO with PLS gives the smaller errors on constrained test cases   but, for unconstrained ones, SEGO-KRG performs the best. As we can see on the data profiles, SEGO-KRG and SEGO+KPLS(d=2) are similar and outperform the three other methods. However, SEGO with PLS being less efficient on unconstrained test cases, SEGO-KRG gives better results over all tests. These preliminary results are promising and some more realistic applications are considered in the next section.

\begin{table}[H]
\vspace*{-0.7cm}
   \caption{Mean errors of each method after 50 iterations.}
   \begin{center}
   \resizebox{\columnwidth}{!}{%
      \begin{tabular}{*{6}{c}}
       \hline
       \textbf{ERRORS} & NSGA2 & Bandit-BO & SEGO+KRG & SEGO+KPLS(d=2) &SEGO+GD  \\
      \hline
      Unconstrained test cases & 14.29\% & 5.84\% &  2.27\% & 5.74\% &  8.41\%  \\
     Constrained test cases &   58.27\% & - &27.61\% & 25.41\% &  47.83\% \\
        \hline
      \end{tabular}
      }
   \end{center}
   \label{tab:meanerror}
\end{table}

\subsection{Aircraft design test cases} 
\label{subsec:res_AD}

For the core MDO application, we apply FAST-OAD on two different aircraft design problems. The first one named ``\texttt{CERAS}'' is a classic well-known aircraft and the second  ``\texttt{DRAGON}'' is a more innovative new one that is currently under development.  
This time, as the evaluations are expensive, we are doing only 10 runs instead of 20. For each run, we draw a random starting DoE of 5 points.  As we have no equality constraint and a small budget, we will not force the constraints to be as large as possible and we will use only the  mean prediction of the constraint models and not the $UTB$ criterion~\cite{SEGO-UTB}. Also, to have realistic results, the constraints violation will be forced to be less than $10^{-3}$. On the following, let MAC denote the Mean Average Chord, VT be the Vertical Tail, HT be the Horizontal Tail and TOFL be the Take-off Field Length. 
 %UTB, MC

\subsubsection{``\texttt{CERAS}'' configuration}

To validate the results of Section~\ref{subsec:res_aca} on a real test case, we are considering the data from the CEntral Reference Aircraft System (``\texttt{CERAS}'') based on an Airbus A320 aircraft. The problem to solve is a constrained optimization problem with 6 continuous design variables, 2 categorical variables and 2 integer ones, for a total of 10 design variables. The presented version of SEGO (with or without the PLS technique) is used as an optimizer in a Multidisciplinary feasible (MDF) architecture where the MDA is computed with FAST-OAD.
The optimization problem is described in~\tabref{tab:ceras} where the total number of variables is reported using the relaxation technique.

\begin{figure}[H]
   \begin{subfigure}[b]{.5\linewidth}
      \centering
	\includegraphics[height=4.5cm]{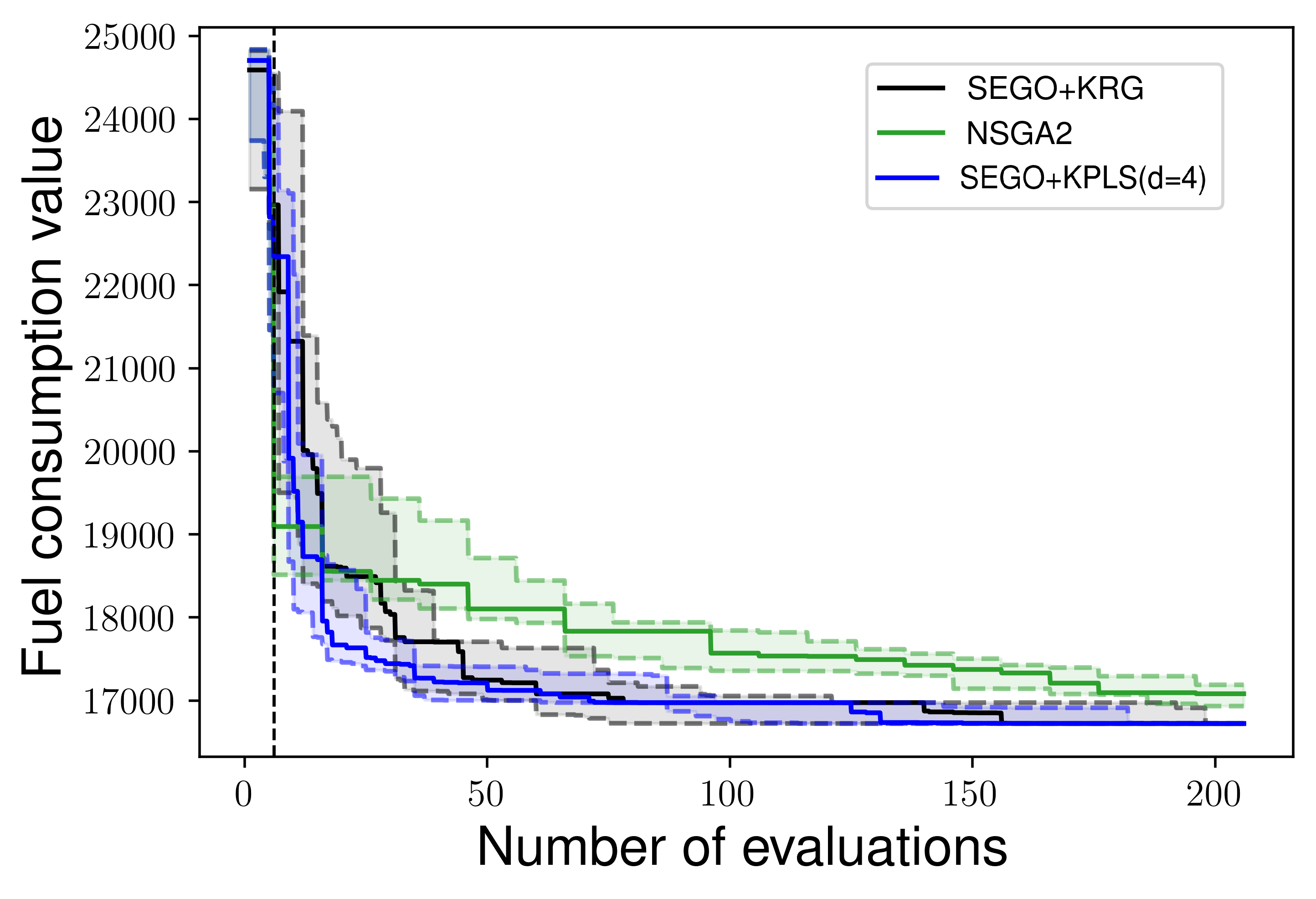}
     \caption{Convergence curves.}
     \label{FASTOADcurves}
      \end{subfigure}
      \begin{subfigure}[b]{.5\linewidth}
      \centering 
\includegraphics[height=4.5cm]{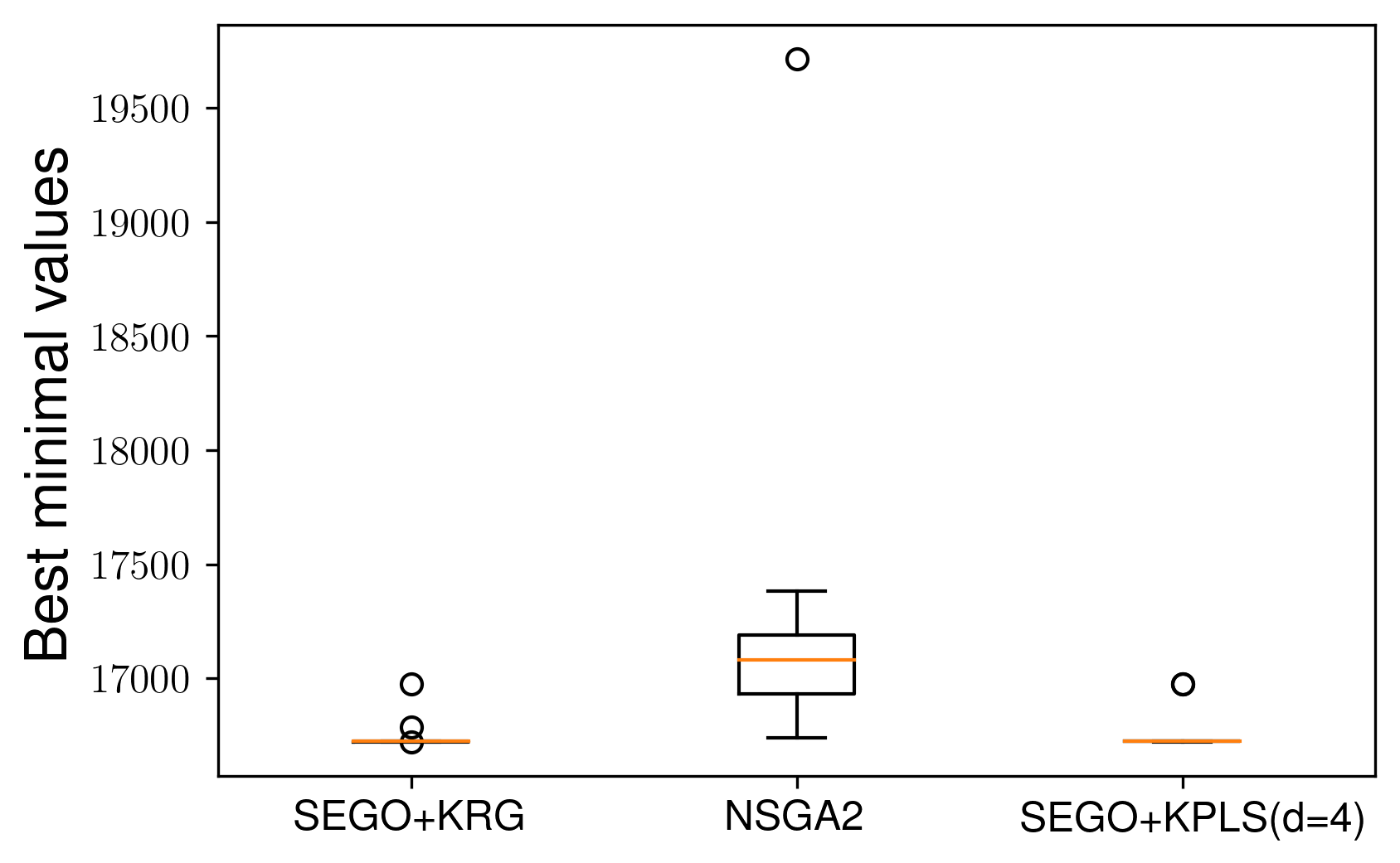}
 \caption{Boxplots.}
 \label{minima_FASToad}
   \end{subfigure}
   \caption{``\texttt{CERAS}'' optimization results using a DoE of 5 points. The Boxplots are generated, after 200 iterations, using the 10 best points.}
   \label{CeRAS}
\end{figure}

\begin{figure}[H]
\vspace*{-0.4cm}
\centering
	\centering	
	\includegraphics[height=5.5cm,width=12cm]{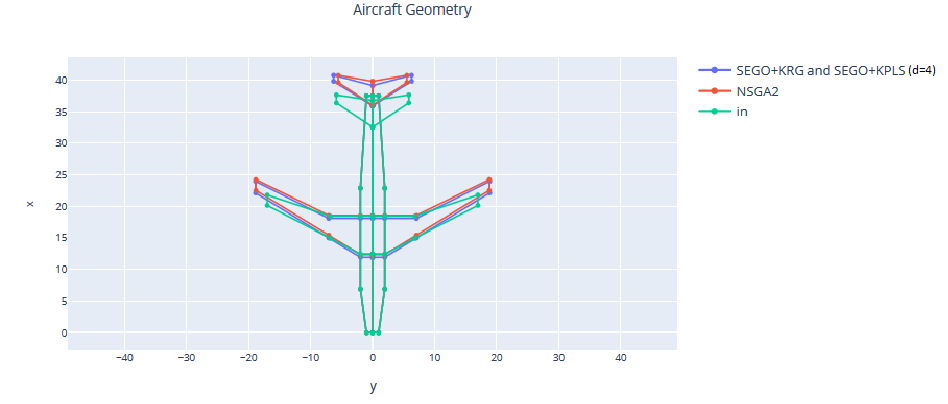}
     \caption{``\texttt{CERAS}'' best configuration geometries. Comparisons between the initial configuration, the NSGA2, the SEGO+KRG and the SEGO+KPLS(d=4) results.}
     \label{CeRASplots}
    \end{figure}

\begin{table}[H]
\vspace*{-0.2cm}
\centering
   \caption{Definition of the ``\texttt{CERAS}'' optimization problem.}
\small
\resizebox{0.9\columnwidth}{!}{%
\small
\begin{tabular}{lclrr}
 & Function/variable & Nature & Quantity & Range\\
\hline
\hline
Minimize & Fuel mass & cont & 1 &\\
\hline
with respect to & \mbox{x position of MAC} & cont & 1 & $\left[16., 18.\right]$ ($m$)\\
 & \mbox{Wing aspect ratio}  & cont &1 & $\left[5., 11.\right]$ \\
 & \mbox{Vertical tail aspect ratio} & cont & 1 & $\left[1.5, 6.\right]$ \\
 & \mbox{Horizontal tail aspect ratio} & cont & 1 & $\left[1.5, 6.\right]$ \\
  & \mbox{Wing taper aspect ratio} & cont & 1 & $\left[0., 1.\right]$ \\
   & \mbox{Angle for swept wing} & cont & 1 & $\left[20., 30.\right]$ ($^\circ$)\\
 & \multicolumn{2}{c}{Total  continuous variables} & 6 & \\
 \cline{2-5}
& \mbox{Cruise altitude} & discrete & 1 & \{30k,32k,34k,36k\} ($ft$)\\
& \mbox{Number of engines} & discrete & 1 & \{2,3,4\} \\
 & \multicolumn{2}{c}{Total  discrete variables} & 2 & \\
 \cline{2-5}
& \mbox{Tail geometry} & cat & 2 levels & \{T-tail, no T-tail\} \\
& \mbox{Engine position}& cat & 2 levels & \{\mbox{front or rear engines}\}  \\
 & \multicolumn{2}{c}{Total  categorical variables} & 2 & \\
 \cline{2-5}
  &   \multicolumn{2}{c}{\textbf{Total relaxed variables}} & {\textbf{12}} & \\
  \hline
subject to & 0.05 \textless \ Static margin \textless \  0.1 & cont & 2 \\
 & \multicolumn{2}{c}{\textbf{Total  constraints}} & {\textbf{2}} & \\
\hline
\end{tabular}
}
\label{tab:ceras}
\end{table}
We are testing three methods, NSGA2, SEGO with Kriging and SEGO with KPLS over 10 different DoE and 200 iterations. As for the analytical test cases,~\figref{FASTOADcurves} shows that PLS method with 4 hyperparameters speeds up the optimization process at the start. However, the boxplots of the final results on~\figref{minima_FASToad} show that, at the end, the KPLS(d=4) and Kriging versions give similar results while KPLS uses 3 times less correlation length hyperparameters. The Kriging version starts slower but catches up with KPLS quickly and for every given budget Kriging gives better results than NSGA2.
The best configuration is obtained after a SEGO+KPLS(d=4) optimization, the optimal result is given in~\tabref{tab:ceras_best}. The optimal aircraft geometries obtained using each of the three methods and the baseline are plotted in~\figref{CeRASplots}. As SEGO with Kriging  is almost the same as the best, SEGO+KRG and SEGO+KPLS are grouped on the geometries. With KPLS we obtain a constraint margin of 0.500 and an objective of 16722.73 kg.

\begin{table}[H]
\vspace*{-0.5cm}
\caption{``\texttt{CERAS}'' Optimal aircraft configuration.}
   \begin{center}
   \resizebox{0.5\columnwidth}{!}{%
\small
      \begin{tabular}{*{3}{c}}
       \hline
        Name & Nature & Value \\
      \hline
      Fuel mass & cont & 16722.55 kg  \\
     static margin &  cont & 0.0495  \\
        \hline
      x position of MAC  & cont &  16.2825 m \\
     Wing aspect ratio  & cont & 11 \\
      VT aspect ratio  & cont & 6 \\
      HT aspect ratio  & cont & 6 \\
     Wing taper ratio  & cont & 0.5099 \\
    Wing sweep angle  & cont &  30.0 $^{\circ}$ \\
    Cruise altitude  & discrete & 36,000 ft \\
     Tail geometry  & cat & T-tail  \\
    Engine position   & cat & Front engine \\ %(Layout 1) 
     Number of engines  & discrete &  2 \\
      \hline
      \end{tabular}
       }
   \end{center}
   \label{tab:ceras_best}
\end{table}
From an aircraft design point of view, the obtained optimized values can be surprising for some parameters, but are logical given the limits of current models in FAST-OAD:
\begin{itemize}
    \item as the flight ceilings (buffeting, aerodynamics, ...) are not computed, the algorithm logically goes toward high altitudes,
    \item the mass of a strongly swept wing is not properly computed, and does not bring enough penalty on such configuration,
    \item similarly, the models do not compute the additional mass that should be required for a T-tail with such a large aspect ratio on vertical tail.
\end{itemize}
However, the capabilities of the proposed algorithm are very promising for a conceptual design stage where a lot of architecture choices are still undetermined, leading to a large combinatorial problem. In that perspective, the ability of the method to capture the right trends regarding number of engines (integer) and their position (categorical) is a good perspective, that needs to be confirmed with an even more complex design case using many categorical and integer variables. The ``\texttt{DRAGON}'' hybrid aircraft case described below provides this feature.  
\subsubsection{``\texttt{DRAGON}'' configuration}

\begin{figure}[H]
\begin{center}
   \begin{subfigure}[b]{.5\linewidth}
      \centering
	\includegraphics[clip=true,height=4cm]{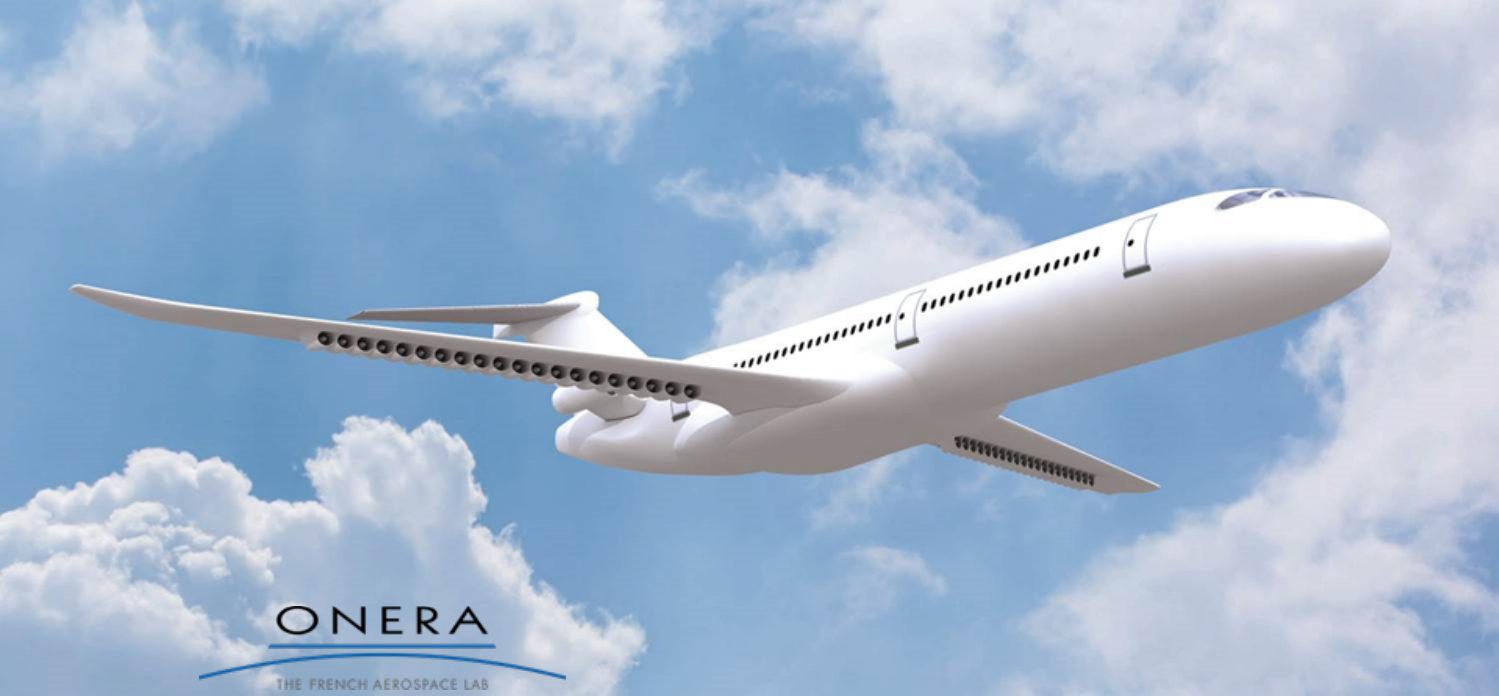}
      \end{subfigure}
\end{center}    
   \begin{subfigure}[b]{.48\linewidth}
      \centering
	\includegraphics[height=5cm]{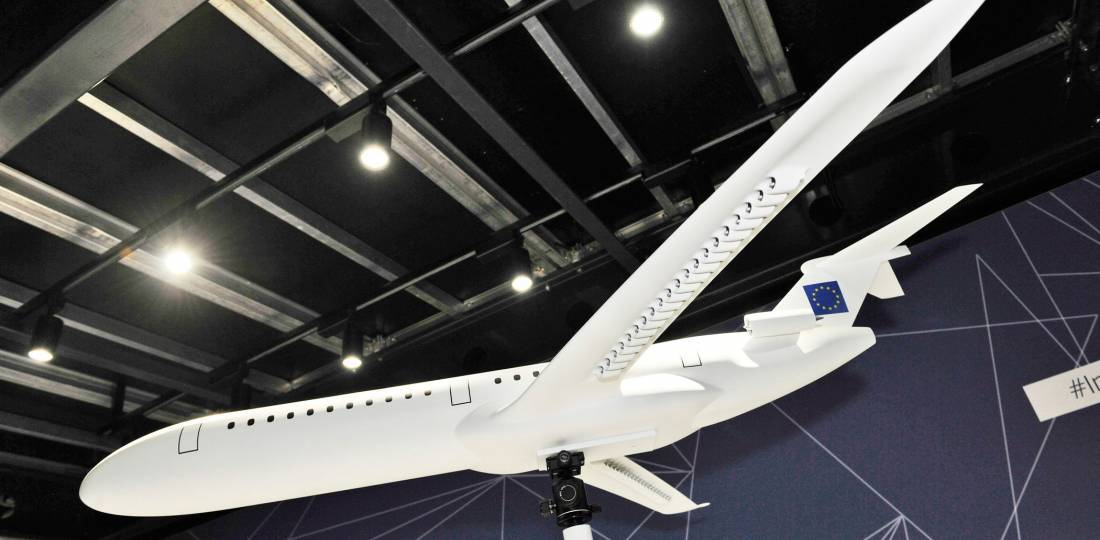}
      \end{subfigure}~~~
      \begin{subfigure}[b]{.48\linewidth}
      \centering 
	\includegraphics[clip=true,height=5cm]{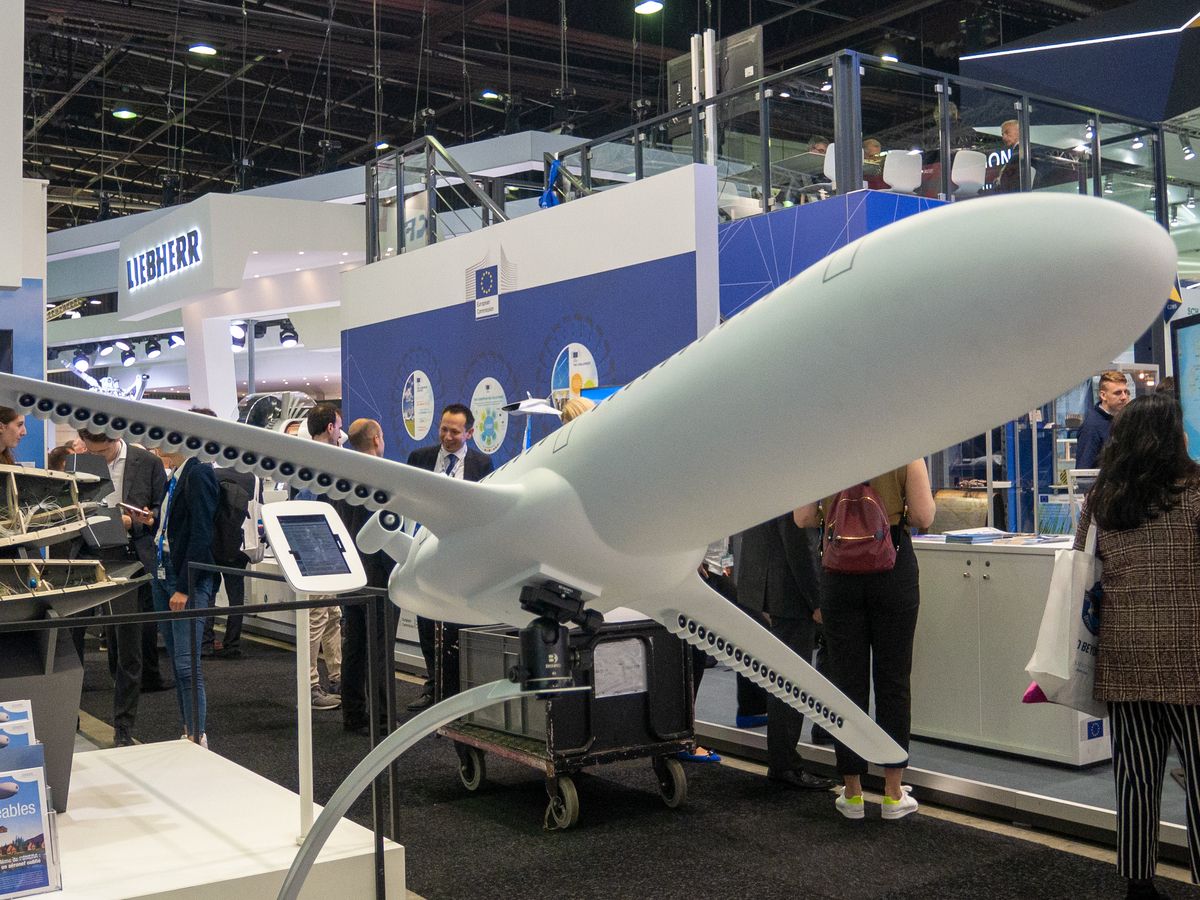}
   \end{subfigure}
   \caption{``\texttt{DRAGON}'' aircraft mock-up.}
  % \label{Dragon2020_IRL}
        \label{Dragon2020}
\end{figure}

The ``\texttt{DRAGON}'' aircraft concept in~\figref{Dragon2020} has been introduced by ONERA in 2019~\cite{schmollgruber} within the scope of the European CleanSky 2 program \footnote{\href{https://www.cleansky.eu/technology-evaluator}{\color{blue}https://www.cleansky.eu/technology-evaluator}} which sets the objective of 30\% reduction of CO2 emissions by 2035 with respect to 2014 state of the art. A first publication in SciTech 2019~\cite{schmollgruber} was followed by an up-to-date report in SciTech 2020~\cite{schmollgruber2}.
ONERA responded to this objective by proposing a distributed electric propulsion aircraft that improves the aircraft fuel consumption essentially by increasing the propulsive efficiency. Such efficiency increase is obtained through improvement of the bypass ratio by distributing a large number of small electric fans on the pressure side on the wing rather than having large diameter turbofans.
This design choice avoids the problems associated with large under-wing turbofans and at the same time allows the aircraft to travel at transonic speed. Thus the design mission set for ``\texttt{DRAGON}'' is 150 passengers over 2750 Nautical Miles at Mach 0.78.

%\begin{figure}[H]
%      \centering
%	\includegraphics[clip=true,height=4cm]{figures/Dragon2020.pdf}
 %    \caption{``\texttt{DRAGON}'' aircraft concept.}
%\label{Dragon2020}
%    \end{figure}

\begin{figure}[H]
    \vspace*{-0.6cm}

      \centering 
	\includegraphics[clip=true,,height=6cm]{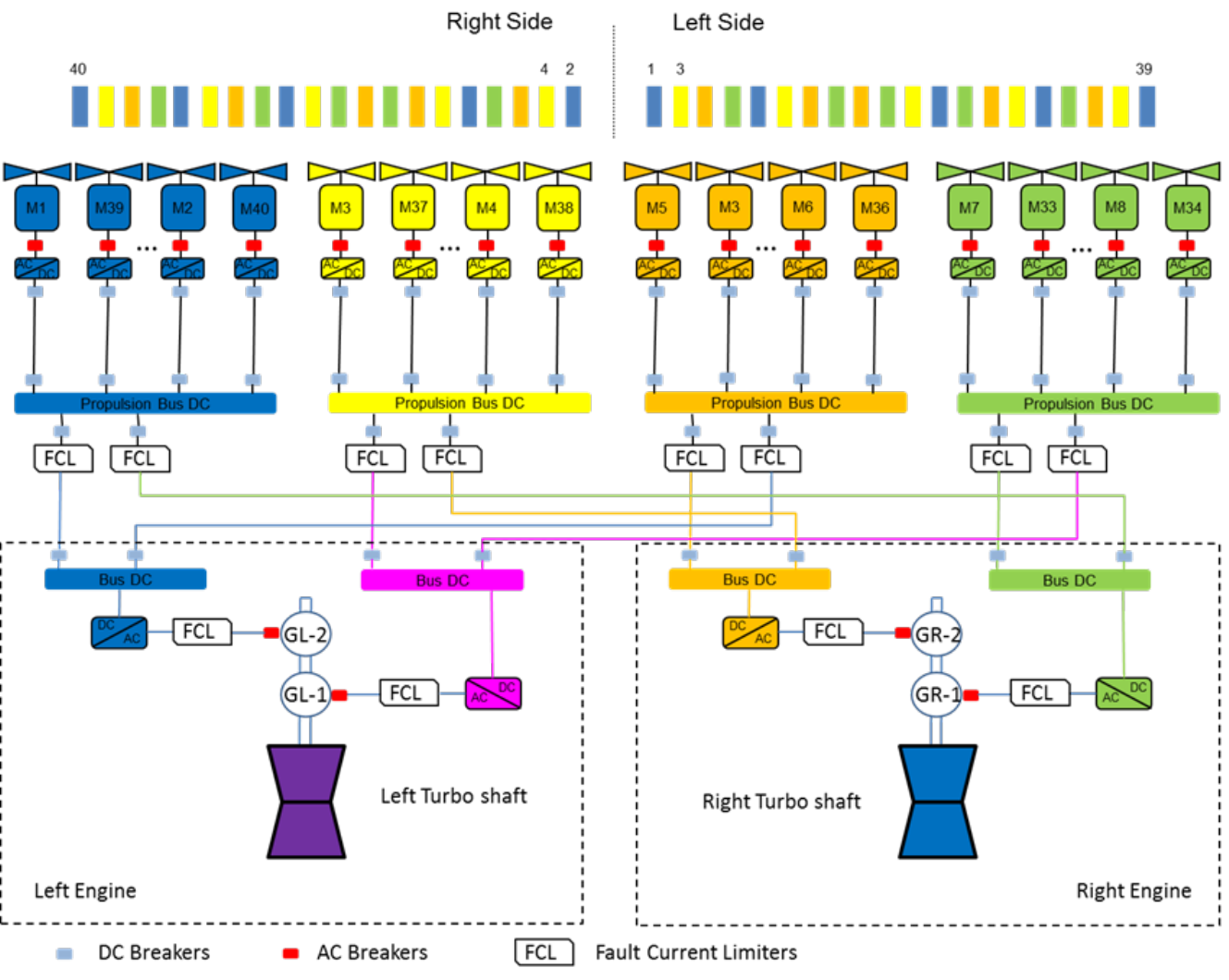}
     \caption{turboelectric propulsive architecture.}
     \label{DragonArchitecture}
   \label{Dragon}
\end{figure}

The employment of a distributed propulsion comes at a certain cost; a turbo-electric propulsive chain is necessary to power the electric fans which brings additional complexity and weight. Typically, turboshafts coupled to electric generators are generating the electrical power on board of the aircraft. The power is then carried to the electric fans through an electric architecture sized to ensure robustness to single component failure. This safety feature is obtained with redundant components as depicted in~\figref{DragonArchitecture}.
The baseline configuration is two turboshafts, four generators, four propulsion buses with cross-feed and forty fans. This configuration was selected for the initial study as it satisfies the safety criterion. However it was not designed to optimize aircraft weight. The turboelectric propulsive chain being an important weight penalty, it is of particular interest to optimize the chain and particularly the number and type of each component, characterized by  some  discrete or particular values.

%\paragraph{The optimization problem} 
In Aerobest~\cite{Aerobest_cat}, we solved a constrained optimization problem with 8 continuous design variables and 4 integer ones, for a total of 12 design variables. Like for “CERAS”, we found that the best configuration with SEGO and Kriging with and without PLS regression were almost the same while PLS reduced the number of variables from 12 to 4. These first results were promising but no categorical variable was considered.
Here the motivation is to add some of these variables in order to increase the number of considered possibilities.
%however the first limitation was the absence of categorical variable which can make the combinatorics explode and is the main motivation of this work. 
%We also had another limitation from the MDA fixed-point algorithm. Now, the DRAGON test case is treated with Overall Aircraft Design method in FAST-OAD~\cite{David_2021} which make it more flexible and general. 
Moreover, the MDA fixed-point algorithm is now replaced to be more flexible and general as the ``\texttt{DRAGON}'' test case is treated with Overall Aircraft Design method in FAST-OAD~\cite{David_2021}.

In~\cite{Aerobest_cat}, the observation of the optimization results from an aircraft design point of view led to surprising conclusions, ultimately resulting in questioning the validity of the models utilized for the design. Among the problems raised in this previous work, the sizing rules of the propulsive chain, based on the default architecture, proved to be insufficiently general to explore more diversified and valid architectures. The results indicated that an architecture comprising 40 motors connected to 6 cores of equal power would be optimal, although it is violating the design rule stating that each motor should be connected to a single core.

In this paper, we propose an up-to-date optimization problem for ``\texttt{DRAGON}'' with the most recent enhancements.
The consideration of the variables related to architecture was revised to take full advantage of mixed variables optimization. Three configurations with different number of electric components were considered, each with their own sizing rules. The default configuration is conserved for the analysis and two new configurations, one with low distribution, the other with high distribution, were created and analysed to establish the sizing rules. The choice of architecture is hence represented by a categorical variable.
The number of motors was to remain an optimization variable as it is an important driver of the propulsive and aerodynamic efficiency of the aircraft but it is constrained by the type of architecture. In this analysis, the number of motors could only be multiple of 4, 8 or 12 depending of the selected architecture. The solution consisted in expending the levels of the categorical variable representing the architecture and assign a specific and valid number of motors to each level.

Additionally, it was noticed that the layout of ``\texttt{DRAGON}'', with the turbogenerators at the rear of the fuselage was disadvantageous for two reasons:
\begin{itemize}
    \item The added weight at the rear forces the wing to be more aft, reducing the level arm with the tail surfaces. To maintain the static stability, the tail surfaces have to be increased, adding weight and friction drag penalty.
    \item The electric cables running from the rear generators to the center part of the wing are heavy, accounting for $\sim$10\% of the total propulsion system weight.
\end{itemize}
For these two reasons, it would be advantageous to locate the turbogenerator below the wing. However, this would restrict the space available for the electric fans, which would in turn limit the maximum propulsive efficiency. To make the trade off between lighter propulsion and better propulsive efficiency the layout was included into the optimization problem as a categorical variable.

Five constraints are added among which, the take off field length, climb duration and top of climb slope angle are strong drivers for the sizing of the hybrid electric propulsion. A portion of the wing trailing edge at the wingtip, should be left free for the ailerons hence limiting the space available for the electric fans. Finally, the wingspan limit is imposed by airport regulation.
%`Parler des améliorations : nouveaux codes de calcul ?
%%Added "WStruct" the version 2 of B. Paluch structure code. Properties:
%%
  %%-GTG below wing or on fuselage
  %%-emotor power requirement at TO, TOC and cruise -> deduce emotor weight and reductor (taken into account so far as fan weight)
%%Modified corresponding modules to accommodate new inputs to WStruct module.
%%Modified cabling after core for the case with 2 cores (each core connected to every motor).

To know how optimizing the fuel mass will impact the aircraft design, we are considering the new optimization problem described in~\tabref{tab:dragon}. We can now solve a constrained optimization problem with 10 continuous design variables and 2 categorical variables with 17 and 2 levels respectively, for a total of 12 design variables. For the optimization, this new problem is a hard test case involving 29 relaxed variables and 5 constraints. The definition of the architecture variable is given in~\tabref{tab:dragon_archi1} and the definition of the turboshaft layout is given in~\tabref{tab:dragon_archi2}.

\begin{table}[H]
\centering
\vspace*{-0.3cm}

 \caption{Definition of the ``\texttt{DRAGON}'' optimization problem.}
\small

\resizebox{0.9\columnwidth}{!}{%
\small

\begin{tabular}{lllrr}
 & Function/variable & Nature & Quantity & Range\\
\hline
\hline
Minimize & Fuel mass & cont & 1 &\\
\hline
with respect to & \mbox{Fan operating pressure ratio} & cont & 1 & $\left[1.05, 1.3\right]$ \\  
     & \mbox{Wing aspect ratio} & cont & 1 &    $\left[8, 12\right]$ \\
    & \mbox{Angle for swept wing} & cont & 1 & $\left[15, 40\right]$  ($^\circ$) \\
     & \mbox{Wing taper ratio} & cont & 1 &    $\left[0.2, 0.5\right]$ \\
     & \mbox{HT aspect ratio} & cont & 1 &    $\left[3, 6\right]$ \\
    & \mbox{Angle for swept HT} & cont & 1 & $\left[20, 40\right]$  ($^\circ$) \\
     & \mbox{HT taper ratio} & cont & 1 &    $\left[0.3, 0.5\right]$ \\
 & \mbox{TOFL for sizing}  & cont &1 & $\left[1800., 2500.\right]$ ($m$) \\
 & \mbox{Top of climb vertical speed for sizing} & cont & 1 & $\left[300., 800.\right]$($ft/min$) \\
 & \mbox{Start of climb slope angle} & cont & 1 & $\left[0.075., 0.15.\right]$($rad$) \\

 & \multicolumn{2}{l}{Total  continuous variables} & 10 & \\
 \cline{2-5}
& \mbox{Architecture} & cat & 17 levels & \{1,2,3, \ldots,15,16,17\} \\
& \mbox{Turboshaft layout} & cat & 2 levels & \{1,2\} \\

 & \multicolumn{2}{l}{Total categorical variables} & 2 & \\
 \cline{2-5}

  &   \multicolumn{2}{l}{\textbf{Total relaxed variables}} & {\textbf{29}} & \\
  \hline
  
subject to & Wing span \textless  \ 36   ($m$)  & cont & 1 \\
 & TOFL \textless  \ 2200 ($m$) & cont & 1 \\
 & Wing trailing edge occupied by fans  \textless  \ 14.4 ($m$) & cont & 1 \\
 & Climb duration \textless  \ 1740 ($s $) & cont & 1 \\
 & Top of climb slope \textgreater \ 0.0108 ($rad$) & cont & 1 \\

 & \multicolumn{2}{l}{\textbf{Total  constraints}} & {\textbf{5}} & \\
\hline
\end{tabular}
}
\label{tab:dragon}
\end{table}

\begin{table}[H]
\centering
\vspace*{-0.3cm}

 \caption{Definition of the architecture variable and its 17 associated levels.}
\small

\resizebox{0.65\columnwidth}{!}{%
\small

\begin{tabular}{cccc}
  Architecture number & number of motors & number of cores & number of generators\\
  \hline
  1 & 8 &2 & 2\\
  2 & 12 & 2 & 2\\
  3 & 16 & 2 & 2\\
  4 &20 &2 & 2\\
  5 & 24 & 2 & 2\\
  6 & 28 & 2 & 2\\
  7 &32 & 2 & 2\\
  8 & 36  & 2 & 2\\
  9 & 40 & 2 & 2\\
 
  10 & 8  &4 & 4\\
  11 & 16 &4 & 4\\
  12 & 24 &4 & 4\\
  13 & 32 &4 & 4\\
  14 & 40 &4 & 4\\

  15 & 12 &6 & 6\\
  16 & 24 &6 & 6\\
  17 & 36 &6 & 6\\

\hline
\end{tabular}
}
\label{tab:dragon_archi1}
\end{table}

\begin{table}[H]
\centering
\vspace*{-0.3cm}

 \caption{Definition of the turboshaft layout variable and its 2 associated levels}
\small

\resizebox{0.7\columnwidth}{!}{%
\small

\begin{tabular}{cccccc}
  layout & position & y ratio & tail & VT aspect ratio & VT taper ratio\\
  \hline 
  1 & under wing &0.25 & without T-tail& 1.8 & 0.3 \\
  2 & behind & 0.34 & with T-tail& 1.2 & 0.85\\
 
\hline
\end{tabular}
}
\label{tab:dragon_archi2}
\end{table}

\begin{figure}[H]
   \begin{subfigure}[b]{.5\linewidth}
      \centering
	\includegraphics[height=4.5cm]{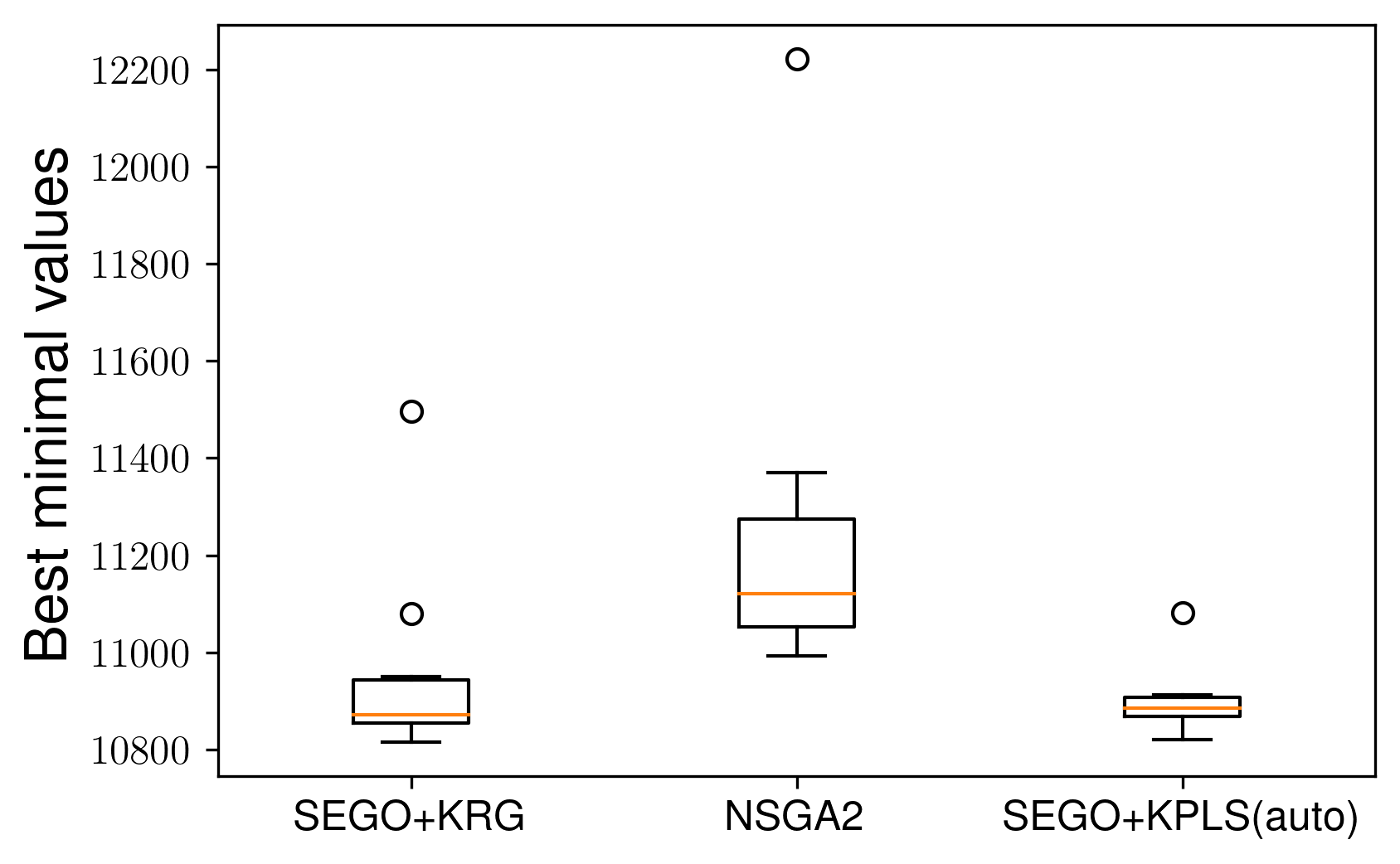}
     \caption{Convergence curves.}
     \label{DRAGONcurves}
      \end{subfigure}
      \begin{subfigure}[b]{.5\linewidth}
      \centering 
\includegraphics[height=4.5cm]{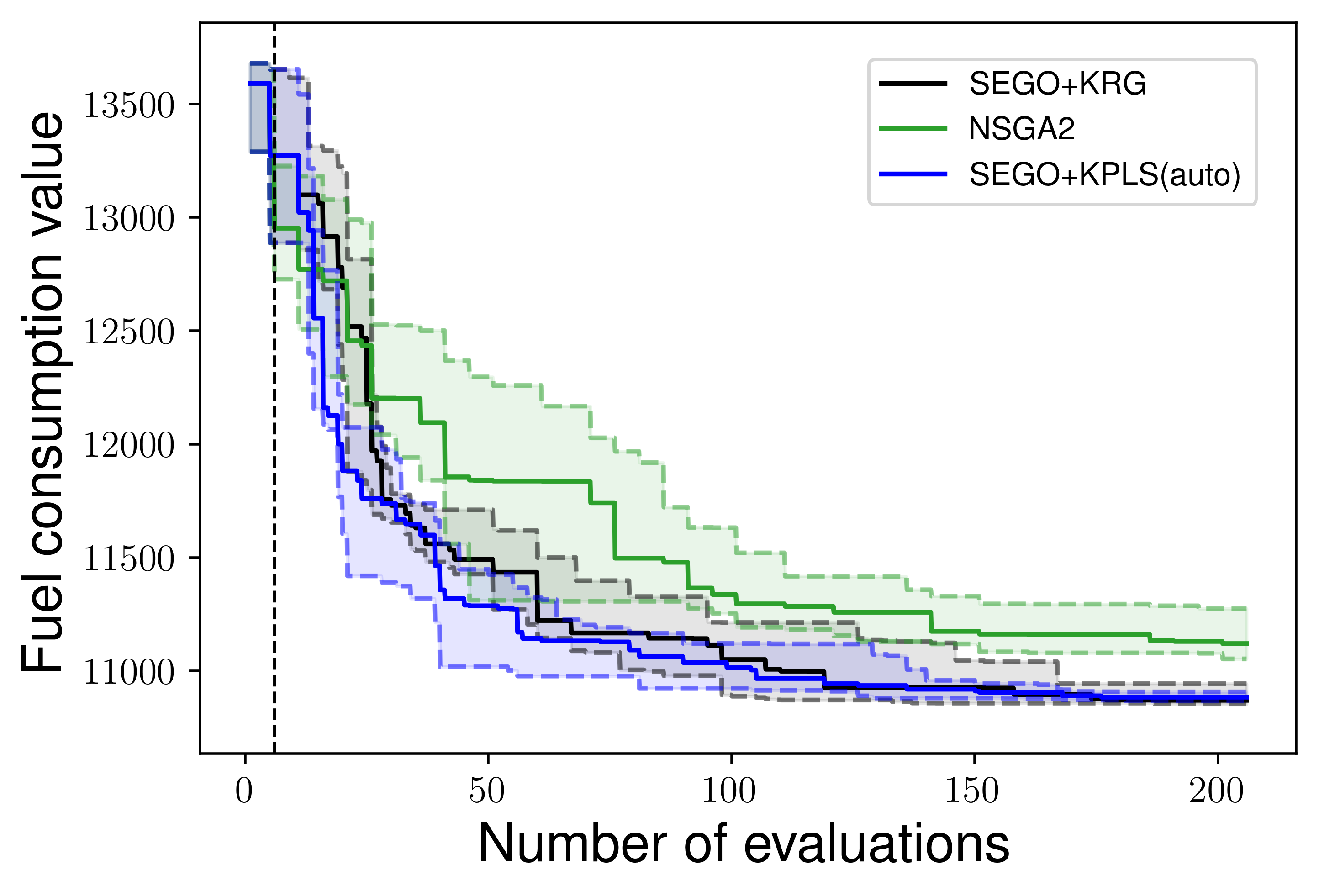}
 \caption{Boxplots.}
 \label{minima_DRAGON}
   \end{subfigure}
   \caption{``\texttt{DRAGON}'' optimization results using a DoE of 5 points over 10 runs. The Boxplots are generated, after 100 iterations, using the 10 best points.}
   \label{DRAGON}
\end{figure}

\begin{figure}[H]
   
      \centering
	\includegraphics[height=5.5cm,width=12cm]{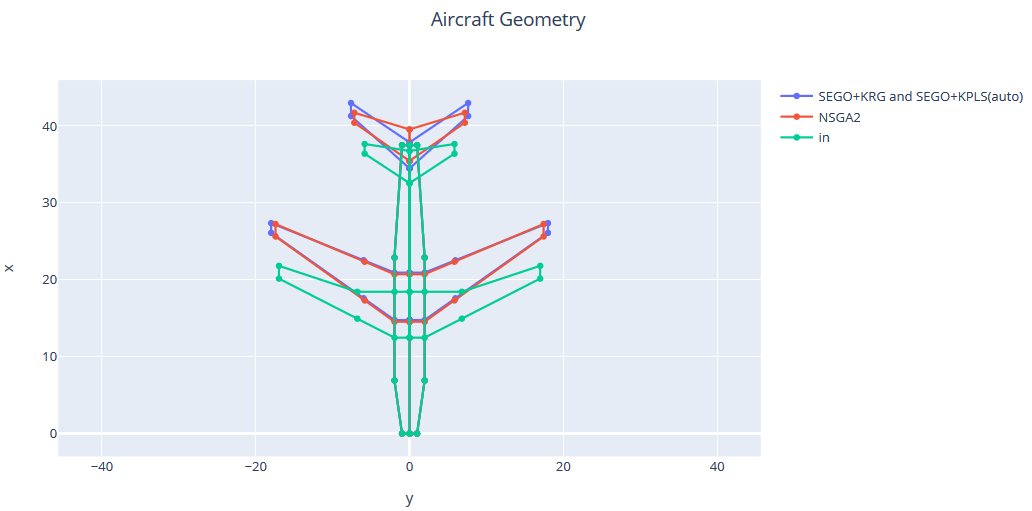}
	\caption{ “\texttt{DRAGON}'' best configuration geometry. Comparisons between the initial configuration, the NSGA2, the SEGO+KRG and the SEGO+KPLS(auto) results.}
     \label{DRAGON_geo}

\end{figure}

To validate our method, we are comparing two variants of SEGO with automatic KPLS or with its more expensive version using Kriging, and NSGA2 in~\figref{DRAGON} where 10 runs are performed in order to plot the mean and the associated quartiles. The best method after 10 iterations is the proposed one involving automatic PLS regression as shown in the boxplots in~\figref{minima_DRAGON}. In~\figref{DRAGONcurves}, after 100 iterations, we still find that SEGO is better than NSGA2 and that the PLS technique helps for the convergence.
As for ``\texttt{CERAS}'', we find that the best configurations found with SEGO and Kriging with and without PLS regression are almost the same. The description of this best configuration is given in~\tabref{tab:dragon_best} and the best geometries for NSGA2, SEGO+KRG and SEGO+KPLS(auto) are given in~\figref{DRAGON_geo}. SEGO+KRG and SEGO+KPLS(auto) are grouped as they give similar results.

\begin{table}[H]
   \vspace*{-0.6cm}

   \caption{``\texttt{DRAGON}'' Optimal aircraft configuration.}
   \begin{center}
   
   \resizebox{0.5\columnwidth}{!}{%
    \small
      \begin{tabular}{*{3}{c}}
       \hline
        Name & Nature & Value \\
      \hline
      Fuel mass & cont &  10816 kg  \\
      Wing span  &  cont & 36 m \\
      TOFL  &  cont & 1722.7 m           \\
      Wing trailing edge occupied by fan &  cont & 10.65 m \\
      Climb duration & cont &  1735.3 s \\ 
       Top of climb slope  &  cont & 0.0108 rad \\
        \hline
     \mbox{Fan operating pressure ratio} & cont & 1.09\\
     \mbox{Wing aspect ratio} & cont & 10.9\\
     \mbox{Angle for swept wing} & cont & 32.2$^\circ$ \\
     \mbox{Wing taper ratio} & cont & 0.235\\
     \mbox{HT aspect ratio} & cont & 6\\
     \mbox{Angle for swept HT} & cont & 40$^\circ$ \\
     \mbox{HT taper ratio} & cont & 0.3 \\
     \mbox{TOFL for sizing} & cont & 1803 m \\
     \mbox{Top of climb vertical speed for sizing} & cont & 494 ft/min \\ 
     \mbox{Start of climb slope angle} & cont & 0.104 rad \\ 
     \mbox{Architecture} & cat & 10 \\
     \mbox{Turboshaft layout} & cat & 2 \\

      \hline
      \end{tabular}
    }
   \end{center}
   \label{tab:dragon_best}
\end{table}

The optimal configuration was found with Kriging for 10816 kg of fuel and KPLS found a similar point at 10819 kg. The best configuration found is the configuration 10, with the smaller number of 8 motors but with 4 cores and electric generators. 

On the point of view of the aircraft design, the number of motors was restricted to the minimum available and the two architectures allowing 8 motors (1 and 10) are performing almost identically, showing a low influence of this variable over the number of motors. In this case, it is the climb constraints that are sizing for the whole propulsion chain. Therefore the architecture sizing laws are not at play and no difference can be seen.

Despite an important space still available at the wing trailing edge, a small number of motors is looked for by the optimizer to lower the wetted surface area of the fans. This is probably oversimplified as the integration of the fans at the trailing edge is coarsely modeled. The fact that this direction is selected during the optimization tells us that this model should be updated to further improve the design.

The most favorable layout is found to be with the turbo-generators at the rear. The level arm between the wing and the horizontal tail being actually larger due to the maximum sweep angle employed for the horizontal tail. As for the “\texttt{CERAS}'' test case, the combination of high sweep and high aspect ratio is too badly taken into account on a structural point of view and leads to unrealistic weight for the horizontal stabiliser, which probably favour too much this layout. Nevertheless, the trade-off found by optimization is correct given the models used in FAST-OAD.

\section{Conclusion and perspectives}
\label{sec:conclu}

To conclude, we have observed on both analytical and industrial cases than SEGO with KPLS is well-suited and efficient for a mixed integer high-dimensional constrained efficient global optimization problem. We have seen that continuous relaxation allows straightforward use of continuous GP but can be impractical as it increases the computational effort required to build the surrogate model. By using the PLS regression, it is possible to reduce the computational cost and makes the continuous relaxation affordable in practical contexts. Using the Wold's R criterion, it is possible to automatize the PLS regression for high-dimensional problems. This method has been applied to aircraft design optimization and contributed to the development of a MDA tool designing future aircraft configurations.

%Future works will involve a new mixed integer Gaussian kernel model for high-dimensional black-box problems to compare with this method and we expect KPLS to be useful when extending to other models. Also, multi-objective optimization will soon be available for aircraft design within SEGO.

%For the Aircraft Design, the MDA for the ``\texttt{DRAGON}'' concept will soon be treated through overall aircraft design within FAST-OAD. The FAST-OAD Tool is still in development and will be made even more flexible and general.

%%%%%%%%%%%%%%%%%%%%%%%%%%%%%%%%%%%%%%%%%%%%%%%%%%%%%%%%%%%%%%%%%%%%%%
\section*{Acknowledgements}

This work is part of the activities of ONERA - ISAE - ENAC joint research group.
The research presented in this paper has been performed in the framework of the AGILE 4.0 project (Towards Cyber-physical Collaborative Aircraft Development) and has received funding from the European Union Horizon 2020 Programme under grant agreement n${^\circ}$ 815122. The authors are grateful to the partners of the AGILE consortium for their contribution and feedback.
We also thanks the people who worked on DRAGON and made this work possible: Carsten Döll, David Donjat, Michael Ridel, Jean Hermetz, Bernard Paluch and Olivier Altinault. 
%%%%%%%%%%%%%%%%%%%%%%%%%%%%%%%%%%%%%%%%%%%%%%%%%%%%%%%%%%%%%%%%%%%%%%

%%%%%%%%%%%%%%%%%%%%%%%%%%%%%%%%%%%%%%%%%%%%%%%%%%%%%%%%%%%%%%%%%%%%%%
% REFERENCES
%%%%%%%%%%%%%%%%%%%%%%%%%%%%%%%%%%%%%%%%%%%%%%%%%%%%%%%%%%%%%%%%%%%%%%

% Produces the bibliography section when processed by BibTeX

\nocite{*}
\bibliography{main.bib}

%%%%%%%%%%%%%%%%%%%%%%%%%%%%%%%%%%%%%%%%%%%%%%%%%%%%%%%%%%%%%%%%%%%%%%
\end{document}